# Passive high-yield seawater desalination at below one sun by modular and low-cost distillation

*Eliodoro Chiavazzo[1,*], Matteo Morciano[1], Francesca Viglino[1], Matteo Fasano[1], Pietro Asinari[1,2,*]*

[1] Department of Energy, Politecnico di Torino, Corso Duca degli Abruzzi 24, Torino 10129, Italy.

[2] Clean Water Center, Politecnico di Torino, Corso Duca degli Abruzzi 24, Torino 10129, Italy.

* Corresponding authors: Eliodoro Chiavazzo, Ph.D.: eliodoro.chiavazzo@polito.it; Pietro Asinari, Ph.D.: pietro.asinari@polito.it

## ABSTRACT

Although seawater is abundant, desalination is energy-intensive and expensive. Using the sun as an energy source is attractive for desalinating seawater; however, the performance of state-of-the-art passive devices is unsatisfactory when operated at less than one sun (<1 kW $m^{-2}$). Here, we present a completely passive, modular, and low-cost solar thermal distiller for seawater desalination. Each distillation stage is made of two opposed hydrophilic layers separated by a hydrophobic microporous membrane, and it does not require further mechanical ancillaries. Under realistic laboratory and outdoor conditions, we obtained a distillate flow rate of almost 3 L $m^{-2}$ $h^{-1}$ from seawater at less than one sun – twice the yield of recent passive device reported in the literature. In perspective, theoretical modelling suggests that the distiller has the potential to further doubling the peak flow rate observed in the current experiments. This layout can satisfy freshwater needs in isolated and impoverished communities, as well as realize self-sufficient floating installations or provide freshwater in emergency conditions.

Four billion people currently face severe water scarcity for at least one month per year, whereas half a billion for all year round. [1-3] This widespread and devastating problem has motivated the development of a large variety of desalination technologies to provide freshwater from the abundant seawater, by means of either membranes or thermal processes. However, most of the systems involve costly or cumbersome solutions. [4-6] Conventional desalination technologies are typically based on active processes, namely, they include components with mechanical moving parts that are subject to aging and possible failure. Since active desalination involves high capital and operating costs, large plants are generally implemented.

By contrast, passive desalination technology is built using self-operating systems, where all processes occur without mechanical moving parts. Although usually less efficient as compared to active technologies, owing to lower capital requirements and operating costs, passive approaches have the potential to improve the economic feasibility and reliability of small plants, especially in isolated and impoverished areas. [7] In particular, solar stills have been known for thousands of years, and they have the major advantage that they function using only the sun. However, even their best realizations can be highly inefficient, and large-area installations are typically required to satisfy the drinkable water needs of a single person: six square meters per person per day. [8]

In recent years, advanced nanomaterials have been used to improve the efficiency of stills by reducing heat losses to the environment and enhancing the efficiency of solar energy absorption. [9-14] Some efforts have been so successful that sunlight could be converted to steam and then condensed with efficiency in the order of 90%, even without optical concentration. Owing to such progress, an incredibly high quantity of distilled water produced by 1 kWh of input solar energy was recently achieved by a solar still made of inexpensive materials, that is 1.28 L $kWh^{-1}$ under one sun (*i.e.*, 1.28 L $m^{-2}$ $h^{-1}$). [15] This milestone can potentially increase the productivity of conventional solar stills by a factor of four, at the cost of a few dollars. [8,15-17] However, while condensing freshwater in a solar still, a large amount of latent energy is always lost into the environment. This occurs regardless of the level of sophistication of the nanomaterials used for optimizing the evaporation process and thermal insulation of the liquid phase. Previous attempts to recover latent heat of condensation to enhance desalination performance intrinsically involved active components (e.g. pumps), as in the case of vertical or inclined multi-effect solar stills; [18-20] whereas, speculations to recover it in a passive way were not supported by experimental evidences, [21] possibly due to distillate contamination issues.

Here, we design, build and test a completely passive solar-driven distiller that is capable of re-using the latent heat of vaporization several times before it is lost to the environment. This process is implemented in a completely passive way at ambient pressure, since the sun provides thermal energy for distillation and capillary forces drive water feeding, which can be used also by floating installations. Furthermore, the latent heat of condensation of distillate is recovered by multiple evaporation/condensation stages, which allow to go beyond the thermodynamic limit of single-stage distiller $q_{solar}/\Delta h_{LV} \cong 1.47$ L m$^{-2}$ h$^{-1}$ under one sun ($q_{solar} = 1$ kW m$^{-2}$), being $\Delta h_{LV} = 2455.6$ kJ kg$^{-1}$ the total enthalpy of liquid–vapour phase change from ambient temperature. [15] In our device, both the passive water feeding and the multistage distillation have been unlocked by the horizontal orientation of distillation stages. However, differently from previous attempts, the innovative use of a hydrophobic membrane between the evaporating and condensing hydrophilic layers helps avoiding distillate contamination under horizontal configuration. We demonstrate, both experimentally and theoretically, that tremendous improvements in terms of the distillate flow rate, ranging from two to four times respect to state-of-the-art passive solar stills, [15] can be easily achieved without compromising simplicity and materials cost. For example, assuming the theoretical distillate flow rate reported in **Supplementary Fig. S10** (namely ≅6 L $m^{-2}$ $h^{-1}$), a single square meter of distiller exposed to direct solar radiation (for 6 working hours) theoretically has the potential to satisfy the daily drinkable water needs (e.g. two litres per day [22]) of up to eighteen people. The reported approach and results represent a remarkable improvement over recently published passive technologies and could possibly have an immediate impact on the life of millions of people in the most impoverished regions of the world.

## RESULTS

**Layout of the passive solar distiller**. Our modular distiller can passively desalinate seawater by exploiting low-temperature heat (here, from non-concentrated solar radiation) without the need for any mechanical or electrical ancillaries, such as pumps or valves, during standard operating conditions. The main elements are shown in **Figs. 1A** and **B,** while further technical details are available in **Supplementary Figs. S1** and **S2**. Each stage of the distiller consists of two highly thermally conductive thin aluminum plates of 12×12 $cm^2$, each one supporting a 1 mm-thick layer of hydrophilic microfibers. The liquid layer separation can be accomplished by either using hydrophobic microporous membranes as commonly done in membrane distillation – MD (for active technologies), [23-25] or using a membrane-free solution by leaving a small air gap between the hydrophilic layers. Both approaches are implemented and tested in this work. We notice though that the use of a hydrophobic membrane can be beneficial for avoiding accidental salt contamination of fresh water on the condensing layer, thus safely allowing to realize submillimetre air gaps and arbitrary orientation of the distiller. Materials for the hydrophilic layer, rigid spacer and hydrophobic microporous membrane are reported in **Supplementary Fig. S3A**, **B** and **C**, respectively.

As depicted in **Fig. 1A**, solar radiation is absorbed and converted into heat on the upper side of the first stage of the distiller. TiNOX®, a commercially available spectrally selective solar absorber, is used to increase the conversion efficiency of the solar radiation into heat. [26] TiNOX® shows high solar absorbance ($\alpha$ = 0.95) and limited infrared emissivity ($\varepsilon$ = 0.04) and thus limited radiative loss at the same time. A transparent thermal insulating layer is adopted at the top surface to reduce convective heat loss between the spectrally selective solar absorber and the ambient environment (see **Supplementary Fig. S3D**). This thermal insulator consists of three 2 mm-thick air layers, which are created between thin films of transparent linear low-density polyethylene (LLDPE) supported by an acrylonitrile butadiene styrene (ABS) frame, manufactured by 3D printing. In these air layers, convective heat transfer is limited and, thus, only thermal conduction occurs. Finally, an aluminium heat sink is placed at the bottom of the distiller to efficiently reject heat from the last stage of the distiller to the ambient environment. It is worth stressing that, in this work, the heat sink purposely operates under natural convection and thus does not require any additional power supply. Note that the heat sink shape was not optimized and, therefore, further gains in efficiency are expected from subsequent device design improvements.

**Working principle of the distiller.** The mechanism underpinning the desalination process of the N-stage distiller is based on reducing the characteristic distance separating the two thin hydrophilic layers, thus enabling an efficient multiple evaporation/condensation process at ambient pressure. In **Fig. 1C** an illustrative stage of the distiller is depicted, where the evaporator and the condenser are the upper and lower hydrophilic layers, respectively. This hydrophilic layer is mainly glued to the aluminium plate, apart from a protruding strip that is immersed in either the saltwater (evaporator) or the distillate (condenser) basin. Water flux to and from the distiller relies only upon capillary forces (owing to the hydrophilicity of the layers and inlet/outlet strips) and gravity, respectively.

Under operating conditions, saltwater rises to the upper hydrophilic layer in each stage (evaporator) because of capillarity. A thermal gradient is generated from the top to the bottom of the distiller by solar radiation on the top surface. Therefore, in each stage, the upper aluminium plate heats the salty water in the evaporator, promoting vapour flux through the air gap or the microporous membrane. As schematically represented in **Fig. 1C**, water in the evaporator and condenser has different vapour pressures because of the temperature and salinity gradients through the stage, which lead to a steady net vapour flux from the evaporating to the condensing hydrophilic layers. In **Fig. 1D** the vapour pressure is plotted as a function of water salinity and temperature, which helps in understanding the above process (see equations (1-3)). Water vapour condensates on each lower hydrophilic layer, where the released latent heat becomes available to drive additional evaporation stages in devices with a multistage configuration. The reuse of heat by subsequent stages is essential for overcoming the performance limitations of current passive distillers. As a result, distilled water accumulates in the hydrophilic condensation layer, and its strip drains the excess freshwater into the basin on the right-hand side (aided by gravity). It is worth pointing out that, contrarily to traditional solar stills where the optical transmittance of the transparent cover may be reduced by condensed water drops, here the condensation process does not affect the optical performance of the device.

**Laboratory experiments.** The passive distiller was tested under both laboratory and outdoor conditions. First, the performance was assessed under laboratory conditions, where a non-fluctuating thermal source from an electrical resistor was adopted to ensure a constant thermal gradient (**Fig. 2B** and **Supplementary Fig. S4**). The distiller prototype was evaluated by the test rig depicted in **Supplementary Fig. S5**. In the laboratory we evaluated distillers with different number of stages (1, 3 or 10 stages) and various condensing-evaporating interface layers (membrane-free, 0.1 μm or 3.0 μm PTFE membranes). The feed water processed in these experiments was a water/NaCl solution that mimics the salinity of seawater at 35 g $L^{-1}$. The distillate fluxes measured for these different distiller configurations are presented in **Fig. 2A** and **Supplementary Tab. S1**.

Varying the number of stages revealed that the 3-stage configuration device gave a three-fold increase in the specific mass flow rate with respect to a 1-stage configuration, and this already goes beyond the thermodynamic theoretical limit of single-stage distillers. Additionally, a six-fold increase relative to a 1-stage configuration was observed using the 10-stage configuration. The experimental distillate flow rates of the passive distillers fall within the error range of theoretical predictions (grey band in **Fig. 2A**, see Methods for details on the theoretical model). In particular, the configuration with 0.1 μm PTFE membranes and 10 stages produced a specific mass flow rate ($J$) of 2.95±0.02 L $m^{-2}$ $h^{-1}$ distillate with salinity <0.01 g $L^{-1}$ (resolution of the refractometer), which corresponds to 3.28±0.04 L $kWh^{-1}$ that is a 2.5 enhancement respect to state-of-the art passive solar desalination systems. [15]

The nonlinear relation between the number of stages and the distillate specific mass flow rate enhancement is due to a reduced temperature difference across each stage, which is gradually less effective at counteracting the vapour pressure gradient imposed by the salinity (see **Figs. 1C** and **D** for a graphical representation of this effect). It is interesting to note that comparison between devices with either air gaps or hydrophobic membranes shows similar performance. While the air gap guarantees a larger temperature gradient at the evaporator-condenser interface because of a lower thermal transmittance (air gap transmittance ≅ 100 W $m^{-2}$ $K^{-1}$; membrane average transmittance ≅ 470 W $m^{-2}$ $K^{-1}$), the permeability coefficient is reduced because of the larger gap thickness (air gap permeability ≅ $8\times10^{-8}$ kg $m^{-2}$ $Pa^{-1}$ $s^{-1}$; membrane average permeability ≅ $6\times10^{-7}$ kg $m^{-2}$ $Pa^{-1}$ $s^{-1}$). This balance between the different heat and mass transfer characteristics of the two solutions results in their similar performance.

**Field experiments.** We next performed experimental tests of the 3- and 10-stage configurations of the distiller under outdoor conditions. The passive distiller was first tested on a rooftop in Torino, Italy. The experimental setup adopted for the outdoor tests is pictured in **Supplementary Fig. S6**. Outdoor measurements were carried out on clear days around noon, when an approximately constant level of solar irradiance, $q_{solar}$ ≅ 600 W $m^{-2}$, was achieved for at least four consecutive hours (see **Fig. 2D** and **Supplementary Figs. S7**). Seawater collected from the Ligurian Sea, with a salinity of 35 g $L^{-1}$, was supplied to the distiller to better match field-testing conditions. When exposed to direct sunlight, the 10-stage distiller achieved 2.07±0.11 L $kWh^{-1}$ distillate productivity, anyway above the thermodynamic limit of single-stage solar stills. In **Fig. 2C** and **Supplementary Tab. S1**, these distillate fluxes are reported in terms of litres of distilled water produced per kWh of solar energy input, for a better comparability between different ambient conditions. The good performance of the distillers in the laboratory was maintained during outdoor tests. By considering outdoor boundary conditions (see **Supplementary Tab.**

**S2**), the rooftop performances of the passive distiller are also within the calculated uncertainties of the theoretical model.

A floating configuration of the 3-stage passive distiller was then tested in the sea. The distiller, when positioned on a floating platform above the sea, is able to pick up seawater, desalinate it by a distillation process that exploits only incoming solar energy, and finally deliver a steady flux of freshwater into a storage basin (see **Supplementary Fig. S8** and **Supplementary Movie S1**). The test was carried out on a clear day from noon for four hours; the solar irradiation was monitored and is depicted in **Supplementary Fig. S9** (850 W $m^{-2}$ on average), and the inlet seawater had a salinity of 35 g $L^{-1}$. In these conditions, the floating distiller showed about 1.77 L $kWh^{-1}$ distillate productivity with salinity $<0.01$ g $L^{-1}$, which is again above the thermodynamic limit of single-stage solar stills and it is in good accordance with modelling predictions (see **Fig. 2C**). Therefore, using a simple proportionality with the yields obtained during rooftop tests, a remarkable distillate productivity of $\approx 3.71$ L $kWh^{-1}$ could be theoretically extrapolated for a 10-stage passive distiller in the sea.

Since experimental results match well with predictions from the theoretical modelling, the model was then used to predict the potential specific mass flux of the distillate as a function of the number of stages and the salinity (see **Supplementary Fig. S10**). We note that the total temperature drop across the distiller might be further increased by optimizing the layered structure of each stage (see **Supplementary Note 2**). In these cases, the distiller has the potential to achieve a specific mass flux of distillate up to 6 L $m^{-2}$ $h^{-1}$ under 900 W $m^{-2}$ (namely $\approx 6.66$ L $kWh^{-1}$) and even to efficiently process feed water with high salinity (e.g. brines).

**Salt removal.** During day hours, the solar-driven distillation process leads to salt accumulation in the hydrophilic layers used as evaporators. This increase in salt concentration progressively reduces the productivity of each distillation stage, because of the lower activity of high-salinity solutions (see equations (1) and (2)). The salinity gradient tends then to vanish during night hours, when distillate production is interrupted while the high-salinity water in the evaporator can diffuse back into the saltwater source (e.g. sea). This salt removal process is also aided by gravity, because of the different density of high- and low-salinity solutions. However, based on our experience, such passive salt removal from the evaporator is a slow phenomenon occurring on longer time scales as compared to the available night hours. Therefore, in the following we describe one possible engineering solution to this problem. The salt evacuation rate given by these passive processes could be enhanced by rinsing the hydrophilic layers with an additional flow of low-salinity water (e.g. 35 g $L^{-1}$). In the configuration detailed in **Fig. 3A**, the hydrophilic layers where evaporation occurs are connected to a rinse basin via additional hydrophilic strips. The rinse basin is empty during day hours ($\Delta z_1 = 0$ cm), while it is filled up with low-salinity water during night hours. The resulting small hydraulic head ($\Delta z_1 \approx 1$ cm) drives the rinse of evaporating layers.

To assess the salt removal efficacy of rinse, the average distillate productivity of a 3-stage distiller was monitored under laboratory conditions over five consecutive days. Each day, the distiller was powered for eight hours by a non-fluctuating thermal source (approximately equivalent to one sun irradiation) to maintain a stable distillate production. Since each stage generated approximately 50 mL of distilled water, ≈1.75 grams of salt accumulated in each evaporating layer after eight hours. During the remaining sixteen hours, the produced distilled water was removed from the distillate basin, while the strips of the three evaporators were kept immersed in the saltwater basin (35 g $L^{-1}$). In case of rinse, each evaporating layer was also connected to the rinse basin through the additional hydrophilic strip. The evaporating layers were then rinsed for one minute with a saltwater flow (35 g $L^{-1}$) equal to 0.8 mL $s^{-1}$, which resulted in an overall amount of 150 mL of saltwater per rinse.

The distillate productivity measured each day was normalized by the value obtained in the first day of experiments, when a new prototype was tested. In **Fig. 3B**, two alternative configurations are compared: salt removal by concentration gradient and gravity (blue squares); salt removal assisted by saltwater rinse (red dots), as sketched in **Fig. 3A**. On the one side, the distillate productivity obtained without rinse showed an exponential decay with time, due to a progressive salt accumulation in the hydrophilic layer of evaporators: after five days, the distiller presented almost a 50% productivity decrease over five days. On the other side, the rinsing process ensured durable distillation performances that, after the first day of test, stabilized around a limited 15% productivity decrease. Note that, to keep the overall desalination process as completely off-grid, a photovoltaic panel should be introduced to power a possible rinse pump. A preliminary energy analysis (**Supplementary Note 3**) indicates that the extension of photovoltaic panel required to power a rinse pump for the 3-stage distiller would be largely lower than 1 $cm^2$, namely less than one-hundredth of the solar absorbing surface currently required by the distiller. In addition, we notice that, regardless of the above rinse, the presence of a pump is requested anyway to move the distillate towards the final user.

## DISCUSSION

Summing up, in this work we presented a passive (*i.e.*, only driven by non-concentrated solar thermal energy), high-yield, modular and low-cost device that is able to desalinate seawater by exploiting the vapour pressure difference across a small gap between two hydrophilic thin layers. The key idea is to use a simple design to reduce the gap between the hydrophilic materials and hence significantly increase the permeability. In our laboratory and outdoor experiments, we observed that this process is efficiently operated by a thermal power density of less than 1 kW $m^{-2}$ and at a maximum temperature of <65 °C. Different methodologies for separating the two hydrophilic layers in each distillation stage were designed and experimentally tested. Furthermore, the desalination performances of the devices with various number of distillation stages were assessed to evaluate the modular design efficiency. The device with the best performance provides a specific mass flux of distillate up to almost 3 L $m^{-2}$ $h^{-1}$ in a 10-stage configuration under laboratory conditions; whereas, theoretical modelling show a potential up to 6 L $m^{-2}$ $h^{-1}$. Such theoretical potential could be experimentally approached by increasing the thermal insulation between distillation stages and ambient, improving the salt removal process, or optimizing the thickness, material and assembly of hydrophilic layers and membranes. Overall, the passive distillers discussed in this work show energy performances that are up to four times higher than that of state-of-the art passive solar desalination systems [9-15] and, although not in the scope of this work, even show some comparable performance as compared to some of the active desalination technologies (see **Fig. 4A** and **Supplementary Tab. S3**).

The low-cost and passive working principle of the distiller introduced in this work may be particularly suitable to provide inexpensive distilled water in case of emergency conditions. For example, in the floating installation tested by experiments (see **Fig. 4B**), seawater is supplied to the modular distiller by capillary action, and it is then desalinated under direct solar energy. These characteristics would be ideal to provide drinkable water after splashdowns, floods, or tsunamis, where off-grid operating conditions are forced temporarily by the emergency condition. The distilled water could also be used to sustain permanently floating gardens (see **Fig. 4C)** for food production and/or $CO_2$ sequestration. It is worth noting that the floating configuration of the distiller improved its performance because of the reduced and stable temperature of the heat sink realized by the sea. The floating configuration of the distiller may also provide freshwater to coastal areas under water stress conditions. For instance, a floating surface of 54 $km^2$ 10-stage distiller at below one sun would be theoretically sufficient to meet the freshwater needs of Rome (240 L per person per day, about 3 million people; see **Supplementary Note 4**). This extension is comparable with the surface of Lake Bracciano (56.5 $km^2$), which has been exploited as freshwater emergency reserve during the recent drought in the Rome area. [27] Note that the modular distiller may be also suitable to treat different feed water types in non-coastal areas, for instance mining, municipal or industrial wastewater. [28-30]

## METHODS

**Experimental materials and methods.** The experimental setup used for evaluating the desalination performance of the passive distiller under laboratory conditions is depicted in **Supplementary Fig. S5**. To mimic the process of solar energy collection, a planar electrical resistor made of enamelled copper wire with 0.4 mm diameter was embedded below the selective solar absorbing layer of the prototype. The electrical resistor was designed to provide a steady heat flux equal to $q = 740$ W $m^{-2}$, which was determined by decreasing the typical peak summer solar irradiation in Torino (approximately 900 W $m^{-2}$) to account for the solar transmittance of the thermal insulator ($\tau \cong 0.86$, transparent LLDPE, measured experimentally) and the solar absorbance of the selective solar absorber ($\alpha \cong 0.95$, TiNOX®). [26] We notice that the latter setup mimics realistic conditions provided that the resistor delivers the expected adsorbed energy (i.e.

for TiNOX® 95% of the incoming sunlight energy), whereas both radiative and convective heat losses are already properly accounted for.

We explored three options for achieving a small separation of the two liquid phases held in the hydrophilic layers in each stage of our distiller. The first option was to maintain an air gap using a ≈1 mm thick polypropylene (PP) spacer with 0.62 porosity (**Supplementary Fig. S3B**). For the second and third options, we used hydrophobic polytetrafluoroethylene (PTFE) membranes with a thickness of 0.15 mm and a pore size of either 0.1 μm or 3.0 μm (**Supplementary Fig. S3C**). When using the hydrophobic membranes, the pore size is critical for avoiding contamination between the liquid phases separated by the hydrophobic membranes, *i.e.,* the smaller the pore size, the higher the liquid entry pressure (LEP) and thus the lower the contamination issues. [31] However, because our system operates at ambient pressure, even large pore sizes are not of particular concern for accidental distillate contamination. In fact, the LEPs of the membranes with pore size 0.1 μm and 3.0 μm are 7.9 bar and 0.25 bar, respectively. The latter value could possibly lead to a faster membrane wetting; however, no salt contamination was observed in the distillate produced during experiments.

The testing facilities adopted under laboratory conditions consist of a laptop for data storage and analysis, a data acquisition board (NI-9213 module for DAQ board, *National Instruments*), a power supply to provide electrical power to the planar electric resistor mimicking field-test conditions, analog (*RS 110 Hanna Instruments*, accuracy ±0.2%) and digital (*HI 96801 Hanna Instruments*, accuracy ±0.2%) refractometers, a precision scale (*Kern PCB 1000-2*, 0.01 g resolution), the distiller prototype and inlet/outlet water basins (100 mL capacity, room temperature). Three thermocouples (*RS Pro*, K-type) connected with the DAQ board recorded the ambient temperature, the average temperature of the spectrally selective solar absorber (evaporator in the first stage), and the average temperature of the heat sink (condenser in the last stage). The latter two temperatures allow the estimate of the overall temperature drop across the N-stage distiller. The laboratory precision scale was used to monitor the mass change over time of the distilled water basin and thus to compute the distillate specific mass flux generated by the distiller. Refractometers were adopted to measure the salt concentration in the saltwater basin as well as to verify the quality of distillate during operation.

The outdoor tests on the rooftop of the Department of Energy at Politecnico di Torino (Torino, Italy) were carried out on 16 and 17 March 2017. Without losing generality, only the evaporator-condenser interface design incorporating a hydrophobic membrane with a pore size of 3 μm was tested. The experimental setups adopted during these outdoor tests (see **Supplementary Fig. S6**) were similar to those used in the laboratory tests, except for the power supply that was no longer needed and a pyranometer (*Delta Ohm LP Pyra 08 BL*) was used to monitor the solar irradiance. In the tests on the rooftop, the distillate flow rates and their uncertainties have been computed from approximately 1 pm to 2:30 pm. The outdoor tests in the sea were carried out in Varazze, Italy on 17 May 2017 (see **Supplementary Fig. S8**), and the distillate flow rate has been computed by considering the cumulative distillate production from approximately 11:30 am to 3:30 pm.

The permeability of the hydrophobic membranes was measured via a diffusion cell (*PermeGear 15 mm Side Bi Side Cell*, 7 mL volume). In this setup, as schematically represented in **Supplementary Fig. S11**, two aqueous salt solutions with different salt concentrations are separated by the clamped hydrophobic membrane. Owing to the concentration gradient, the generated osmotic pressure promotes water flux through the membrane. Freshwater flows through the membrane to the cell containing the solution with higher salinity, and the mass flow rate (thus the membrane permeability) can be evaluated by monitoring the water level in a graduated column. [32] Both sides of the diffusion cells are agitated by stir bars to avoid possible concentration polarization issues at the membrane interfaces.

**Theoretical model.** The driving force for the N-stage distillation process is the difference in water vapour pressure between the evaporating (E) and condensing (C) hydrophilic layers due to both temperature and salinity differences through the air gap or the hydrophobic membrane. [31,33] The vapour pressure gradient can be computed using Raoult's law as follows:

$$\Delta p_v = a(Y_E)p_v(T_E) - a(Y_C)p_v(T_C), \tag{1}$$

where $a$ denotes the activity of water; $Y_E$ and $Y_C$ are the mass fractions ($Y = m_{salt}/m_{solution}$) of salt in the feed and distilled solution, respectively; $p_v$ is the water vapour pressure; and $T_E$ and $T_C$ are the temperatures of the feed and distilled solutions, respectively. [34] Under ideal conditions, the activity of a NaCl aqueous solution can be estimated as:

$$a = \frac{M_{NaCl}(1-Y)}{M_{NaCl}(1-Y) + N_{ion}\, M_{H_2O}Y}, \tag{2}$$

where $N_{ion} = 2$ for NaCl and $M_{NaCl}$ and $M_{H_2O}$ are the molar masses in grams per mole of sodium chloride and water, respectively. The feed water processed in the experiments has a salinity of 35 g $L^{-1}$ ($Y_E = 0.035$), typical for seawater; therefore, equation (2) predicts $a(Y_E) \cong 0.98$. The activity of the distillate is equal to 1, the same as distilled water. The vapour pressure can be evaluated via Antoine's semi-empirical correlation:

$$\log p_v = A - \frac{B}{C+T}, \tag{3}$$

where $p_v$ is intended in mmHg, $A$, $B$ and $C$ are material-specific constants in this case equal to 8.07, 1730.63 and 233.42, respectively. [35] In **Fig. 1D**, the vapour pressure is plotted as a function of water salinity and temperature, according to equations (1–3). The operating conditions of an illustrative single stage of the distiller are reported, where saltwater (35 g $L^{-1}$ salinity) in the evaporator and distilled water in the condenser have 55 °C and 45 °C temperature, respectively. It is worth pointing out that, for the sake of illustration, we are neglecting the thermal resistances of the hydrophilic layers. Under these conditions, **Fig. 1D** illustrates both $\Delta p_v$ during operating conditions ($\Delta T = 10$ °C), and the minimum temperature drop ($\Delta T_{min}$) needed to carry out the distillation process.

Finally, the resulting specific mass flow rate of the distillate ($J$, kg $s^{-1}$ $m^{-2}$) is proportional to the partial pressure gradient via the permeability coefficient ($K$) of the gap between the hydrophilic layers:

$$J = K\Delta p_w. \qquad (4)$$

In each stage of the distiller, the overall permeability coefficient ($K$) reported in equation (4) can be estimated as a series of contributions from the membrane and spacer permeability, [31] namely

$$\frac{1}{K} = \frac{1}{\frac{\epsilon_m PDM_{H_2O}}{\tau RTd_m P_a}} + \frac{1}{\frac{2M_{H_2O}\epsilon_m r}{3RTd_m\tau}\sqrt{\frac{8RT}{\pi M_{H_2O}}}} + \frac{1}{\frac{\epsilon_s PDM_{H_2O}}{RTd_a P_a}}, \qquad (5)$$

where $\epsilon_m$ is the porosity of the membrane, $P$ is the total pressure (vapour + air), $P_a$ is the partial pressure of air, $D$ is the vapour diffusion coefficient of water vapour in air at the mean stage temperature, $M_{H_2O}$ is the molar mass of water expressed in grams per mole, $\tau$ is the tortuosity factor of the membrane, $R$ is the gas constant (8.314 J $K^{-1}$ $mol^{-1}$), $T$ is the mean stage temperature, $\epsilon_s$ is the porosity of the spacer – if used (see **Supplementary Fig. S3B**), $r$ is the average pore diameter, and $d_m$ and $d_a$ are the membrane and spacer thicknesses, respectively. Note that it is possible to empirically estimate $PD = 1.19 \times 10^{-4}\, T^{1.75}$ (expressed as [Pa $m^2$ $s^-$ [1]]). [36] Clearly, $\frac{1}{\frac{\epsilon_m P D M_{H_2O}}{\tau R T d_m P_a}} = 0$ and $\epsilon_s = 1$ in the case of a simple air gap between the hydrophilic layers. The correlation between tortuosity and porosity of the membrane is modelled by the Mackie–Meares equation as: [37-39]

$$\tau = \frac{(2 - \epsilon_m)^2}{\epsilon_m}. \quad (6)$$

Note that, for the membranes with 0.1 μm average pore size, the transport resistances in the membrane are dominated by collisions between pore walls and vapour molecules and, thus, only Knudsen transport resistance takes place. [40]

As shown in equation (5), the overall gap thickness of each distiller stage ($d_g$, which accounts for the air gap, spacer and/or membrane thicknesses) is crucial and strongly affects the permeate flux. To achieve high permeability and thus high mass flux, a balance between minimizing heat transport and maximizing mass transport between the layers should be found. [31,33] The estimates of the membrane permeability by equation (5) are in good agreement (see **Supplementary Fig. S12**) with those measured experimentally by the diffusion cell in **Supplementary Fig. S11**.

In each stage of the distiller, the heat flux ($Q$) between the evaporating and the condensing hydrophilic layers is mainly due to water phase changes and heat transfer by conduction, namely,

$$Q = \frac{k_{eff,g}}{d_g}(T_E - T_C) + J\Delta H_v + Q_l, \quad (7)$$

where $k_{eff,g}$ is the effective thermal conductivity in the gap, including conduction through the air, spacer and/or membrane; $T_E$ and $T_C$ are the mean temperatures of the feed and permeate solution, respectively; $J$ is the specific mass flow rate of water through the gap; $\Delta H_v$ is the latent heat of vaporization; and $Q_l$ is the heat loss through the lateral surface of the stage. An equivalent thermal resistance circuit that schematically represents this one-dimensional thermal model is presented in **Supplementary Fig. S13**.

Note that each 1 mm-thick synthetic microfiber hydrophilic layer was glued (a few drops of silicone adhesive) to a 1.3 mm thick aluminium plate to allow evaporator and condenser assembly in each stage. The global heat transfer coefficient of this assembly (synthetic microfiber + aluminium plate) was experimentally measured as ≅ 250 W $m^{-2}$ $K^{-1}$. Furthermore, the thermal conductivity of the wet synthetic microfiber was estimated as a weighted average between the thermal conductivity of the dry synthetic microfiber (0.04 W $m^{-1}$ $K^{-1}$) and water (0.6 W $m^{-1}$ $K^{-1}$), where the quantity of absorbed water can be estimated by comparative weight measurements between dry and wet hydrophilic layers.

**Statistical analysis.** A sampling period of 15 minutes was used for mass flow rate computation. Experimental measures were performed when a steady self-sustained mass flux of output distillate was achieved, and tests continued for up to four consecutive hours to assess the stability of the distillation process. Standard deviations of measured distillate flow rate were computed from the analysis of the time series (15 min each) of experiments at the steady state. In the plots, error intervals are reported in terms of ± one standard deviation.

The errors in the model predictions are calculated from the uncertainties involved in the measurement of the membrane porosity, the convection coefficients and the assembly geometry (see **Supplementary Tab. S2**). In fact, the non-homogeneous layered structure of the stages generates non-ideal contacts between adjacent layers. Therefore, an additional air gap between the membranes and hydrophilic layers is considered in the model.

The authors declare that the data supporting the findings of this study are available within the paper and its Supplementary Information.

## ACKNOWLEDGMENTS

The authors are grateful to the NANOSTEP (La Ricerca dei Talenti, Fondazione CRT – Torino) project. E.C. acknowledges partial financial support by the Politecnico di Torino through the Starting Grant 56_RIL16CHE01. The authors also thank Maurizio Bressan and Rocco Costantino for laboratory support, Silvia Pezzana (Almeco Group) for providing experimental materials, and Piero Ornolio, Marina di Varazze and Lega Navale Varazze for hosting field tests of the floating distiller.

## AUTHOR CONTRIBUTIONS

P.A. initially suggested the study of a passive floating panel exploiting materials with high solar absorptance and hydrophobic microporous membranes for seawater desalination. E.C. conceived both the idea of thin hydrophilic layers separated by an air gap as a passive distiller unit (membrane-free solution) as well as the multistage idea to gain freshwater flux. P.A. developed the theoretical model and suggested the rinse strategy. M.M. and M.F. assembled the prototypes and conducted computations. M.M. with the help of F.V. and M.F. conducted the experiments. E.C. and P.A. with the help of M.F. supervised the research. All authors contributed to writing the paper.

## ADDITIONAL INFORMATION

The authors declare no competing financial interests. Details and further experimental results are discussed in the Supplementary Information. Correspondence and requests for materials should be addressed to E.C. and P.A.

## FIGURES

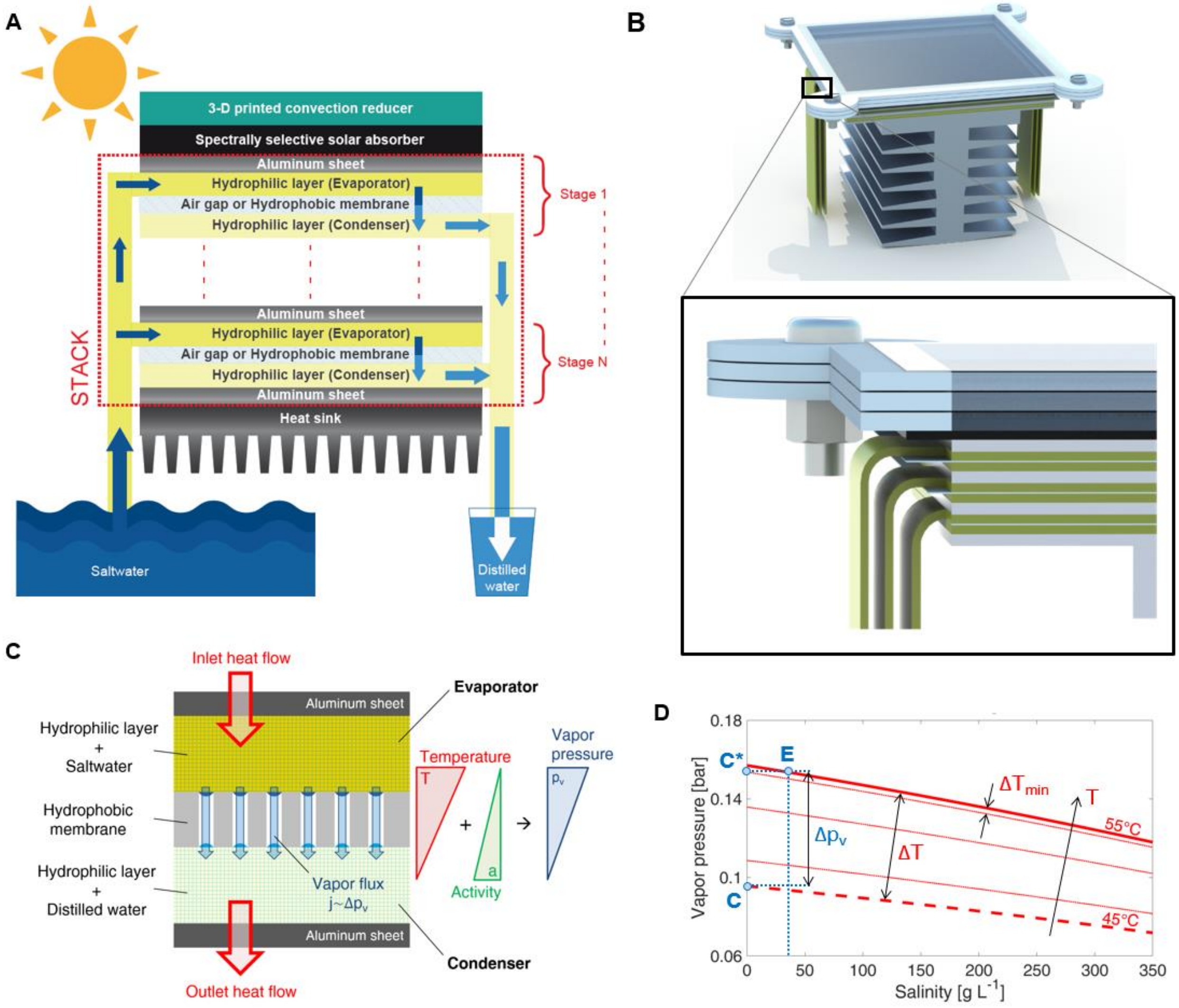

**Figure 1**. **Layout of the solar passive distiller during day hours. (A)** Schematic of the N-stage configuration of the modular distiller. The working principle is (i) seawater is provided to the device by capillary action through strips protruding from the hydrophilic layers of each stage, (ii) solar radiation is converted into thermal energy by a spectrally selective material and is then employed to drive multiple evaporation and condensation stages through either air gaps or hydrophobic membranes, and (iii) distilled water is finally collected in a basin by gravity force. A 3D-printed transparent insulator and an aluminium heat sink were adopted to either reduce or enhance convective heat transfer at the top or bottom surfaces of the distiller, respectively. **(B)** Isometric view and section detail of the 3-stage solar desalination prototype. Note that the yellow, hydrophilic strips are then immersed in either saltwater (evaporators) or distilled (condensers) water basins. During experiments, input and output strips are not directly exposed to air: each strip is covered by a LLDPE film to suppress natural evaporation. Dimensions of components are reported in **Supplementary Fig. S1**. **(C)** Working principle of a single stage of the distiller. The vapour flux from evaporation (saltwater) to condensation (distillate) layer is proportional to the vapour pressure gradient through the stage, which results from the contrasting effects of temperature and activity (*i.e.*, salinity) gradients through the hydrophobic membrane. **(D)** Vapour pressure as a function of water salinity and temperature, where the set of red lines are isothermal curves. The red solid and dashed lines are the isothermal curves of the evaporator (55 °C) and the condenser (45 °C) in the 1-stage distiller reported in **Supplementary Fig. S4A**, respectively. During operating conditions, the temperature gradient (in this schematic represented neglecting the thermal resistances of hydrophilic layers and air) $\Delta T = 10$ °C induces a vapour pressure gradient $\Delta p_v$ between evaporator (point E, saltwater with 35 g $L^{-1}$) and condenser (point C, distillate with 0 g $L^{-1}$), and thus a net vapour flux. The point $C^*$ represents the state of the condensing layer with the minimum temperature gradient $\Delta T_{min}$ to carry out water distillation.

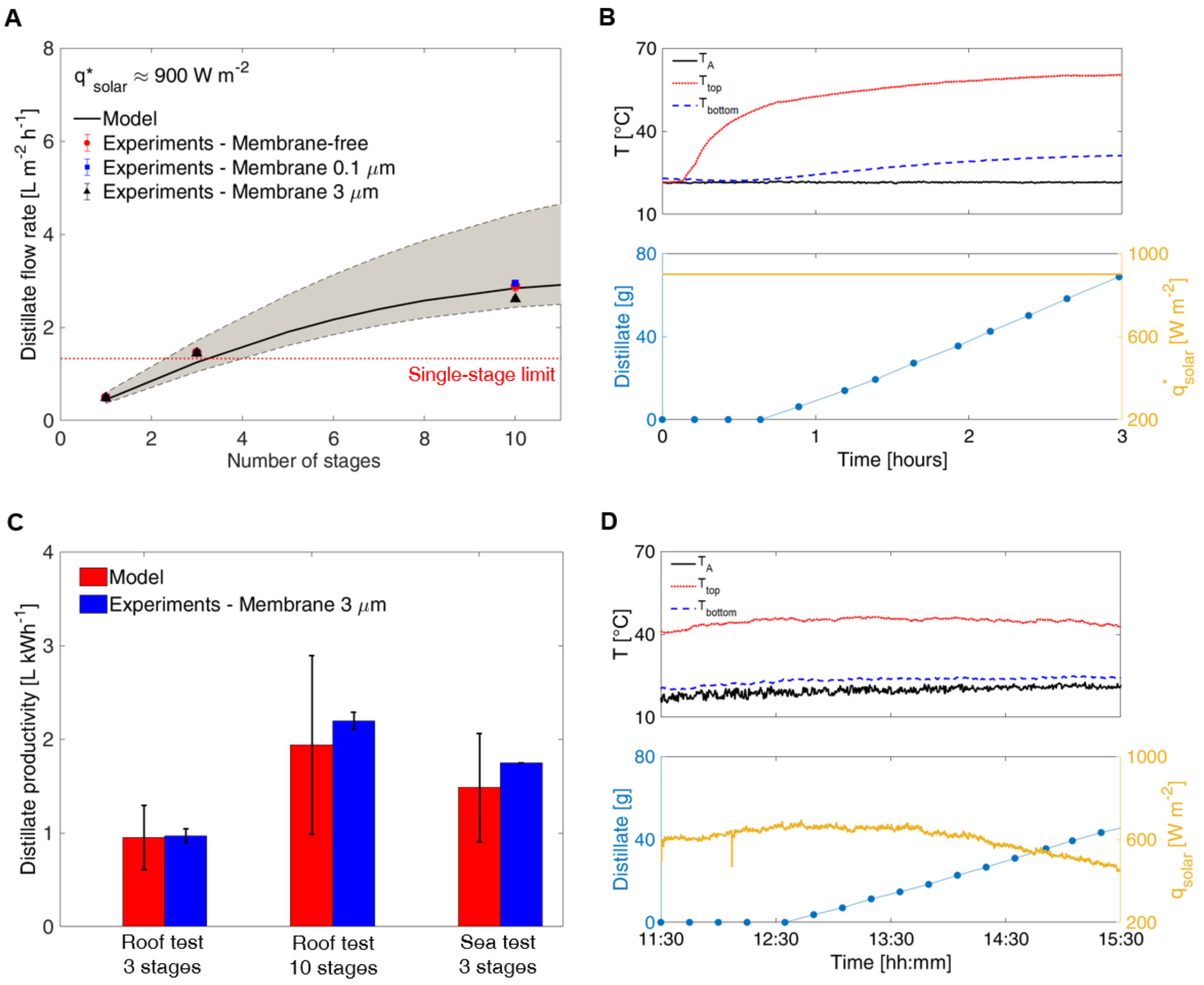

**Figure 2. Desalination performance of the modular distiller. (A)** The desalination performance of the modular distillers were tested under laboratory conditions, with 740 W m$^{-2}$ input thermal energy (about $q^*_{solar}$ = 900 W m$^{-2}$ equivalent solar irradiance which includes the solar transmittance of the thermal insulator $\tau$ and the solar absorbance of the selective solar absorber $\alpha$) and a different number of stages. The distiller was tested with different interface configurations between evaporation and condensation layers: membrane-free, a hydrophobic membrane with 0.1 µm pore size, and a hydrophobic membrane with 3.0 µm pore size. The model prediction and uncertainty for the membrane with 3.0 µm pore size are represented by the black line and the grey band, respectively. The red dotted line reports the thermodynamic

theoretical limit of single-stage distillers under 900 W $m^{-2}$ (1.32 L $m^{-2}$ $h^{-1}$). See **Supplementary Note 1** for details. **(B)** Temperature profiles, distillate production, and equivalent solar irradiance during the laboratory test of the 10-stage distiller. Solid black, dotted red, and dashed blue lines represent the ambient ($T_A$), top first stage evaporator ($T_{top}$), and bottom last stage condenser temperatures ($T_{bottom}$), respectively. **(C)** Theoretical and experimental desalination performances of passive distillers with a different number of stages, under outdoor conditions. The distillate productivity indicates the litres of distilled water produced per kWh of solar energy input. Roof and sea tests refer to the configurations depicted in **Supplementary Figs. S6** and **S8**, respectively. **(D)** Temperature profiles, distillate production, and solar irradiance during the outdoor test (roof, 17 March 2017, 45°03'43.0"N 7°39'35.7"E Torino - Italy) of the 10-stage distiller. Solid black, dotted red, and dashed blue lines represent the ambient ($T_A$), top first stage evaporator ($T_{top}$), and bottom last stage condenser temperatures ($T_{bottom}$), respectively. The average distillate flow rate and its uncertainty have been computed in the interval from 13:00 to 14:30.

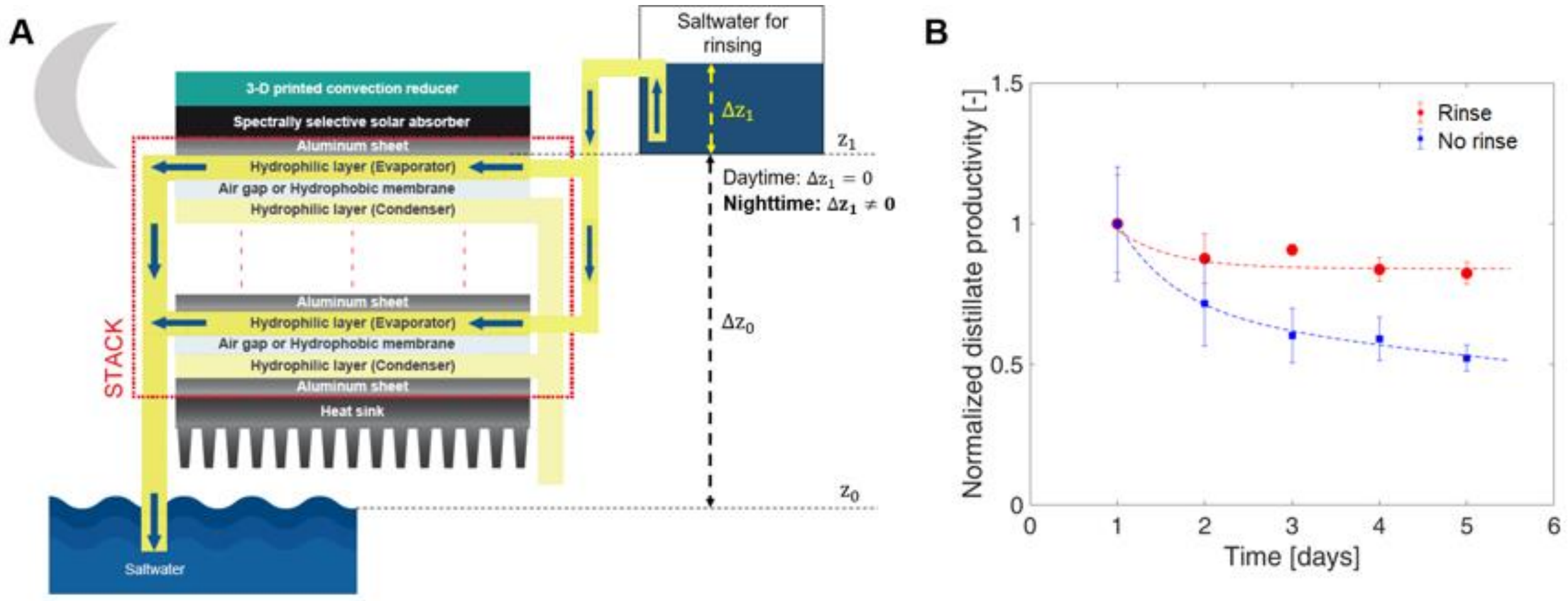

**Figure 3**. **Salt removal and durability of desalination performance during night hours. (A)** The solar-driven distiller operates during day hours, therefore progressively increasing the salt concentration in the evaporating layers. During night hours, instead, the salt accumulated in the evaporators diffuses back into the saltwater basin due to concentration gradient and gravity. This salt removal process can be improved by rinsing the hydrophilic layers of evaporators with a saltwater flow driven by hydraulic head ($\Delta z_1$). **(B)** Durability of desalination performance in a 3-stage distiller under laboratory conditions. The distillate productivity measured each day (L kWh$^{-1}$) is normalized by the value obtained in the first day of experiments, when a new prototype was tested. Day (distillate production, salt accumulation) and night (no distillate production, salt removal) operations are considered for five consecutive days. Two alternative configurations are compared for the night operations: salt removal by concentration gradient and gravity (blue squares); salt removal assisted by saltwater rinse (red dots). Red and blue lines are guides for the eyes.

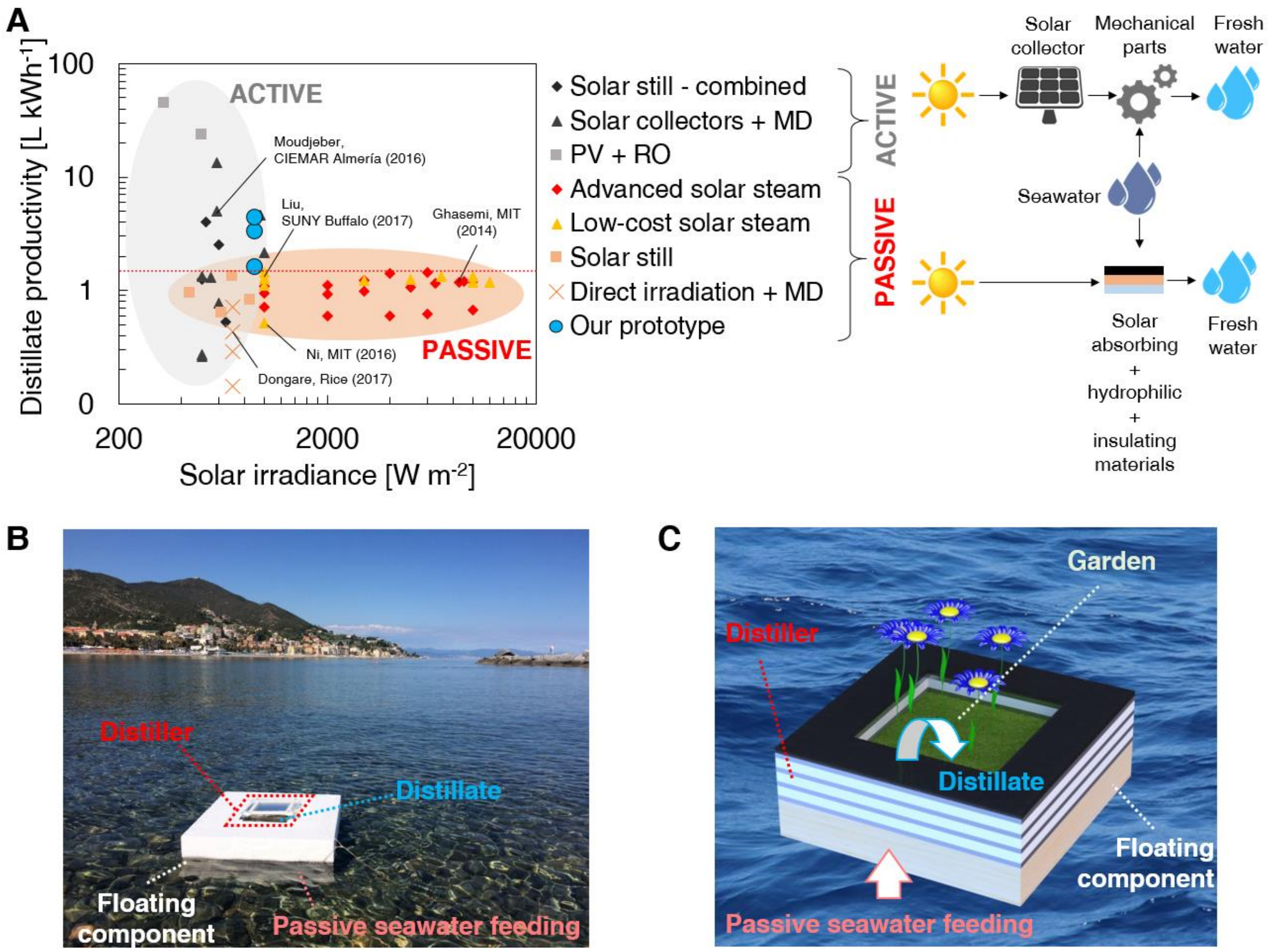

**Figure 4**. **Comparison and applications of the modular passive distiller. (A)** Comparison between the energy desalination performances (*i.e.,* litres of distilled water produced per kWh of solar energy input) of active and passive solar desalination technologies in the literature. Unlike active devices that include mechanical moving parts, the working principle of passive technologies only relies on combinations of solar-absorbing, hydrophilic layers and thermally insulating materials. Technology based on distillation, membrane distillation (MD), and reverse osmosis (RO, coupled with photovoltaics – PV) processes are depicted. Black rhombi indicate solar stills combined with active components. The performances of the distiller discussed in this work are represented by blue dots, whereas the complete list of results reported in the picture is given in **Supplementary Tab. S3**. The highlighted results refer to the works of Moudjeber *et al.*

[41], Liu *et al.* [15], Ghasemi *et al.* [9], Dongare et al. [42], and Ni *et al.* [13]. The red dashed line reports the thermodynamic limit of single-stage distillers under one sun (1.47 L $kWh^{-1}$). **(B)** Floating installation of the modular distiller, which could be employed in emergency conditions (e.g. splashdown, flood, tsunami). A larger amount of drinkable water could be achieved by mosaic-like arrangements of small-sized distillers, each one fed by separated hydrophilic strips in order to limit and optimize the water transport distance by capillarity. **(C)** Possible configuration of the floating desalination device tested in this work, where the distilled water could be used to sustain floating gardens.

# Passive high-yield seawater desalination at below one sun by modular and low-cost distillation

# - Supplementary Information -

*Eliodoro Chiavazzo[1,*], Matteo Morciano[1], Francesca Viglino[1], Matteo Fasano[1], Pietro Asinari[1,2,*]*

[1] Department of Energy, Politecnico di Torino, Corso Duca degli Abruzzi 24, Torino 10129, Italy.

[2] Clean Water Center, Politecnico di Torino, Corso Duca degli Abruzzi 24, Torino 10129, Italy.

* Corresponding authors: Eliodoro Chiavazzo, Ph.D.: eliodoro.chiavazzo@polito.it; Pietro Asinari, Ph.D.: pietro.asinari@polito.it

## Supplementary Notes

**Supplementary Note 1. Details about laboratory performance of distiller.** Under laboratory conditions (see **Supplementary Fig. S5**), the membrane-free configuration produced a specific mass flow rate ($J$) of 0.498±0.023, 1.462±0.050, and 2.892±0.124 L $m^{-2}$ $h^{-1}$ with 1, 3 and 10 stages, respectively (red circles in **Fig. 2A**). On the other hand, the configuration with 0.1 μm pore size hydrophobic membranes produced $J$ = 0.481±0.007, 1.442±0.040, and 2.947±0.024 L $m^{-2}$ $h^{-1}$ with 1, 3 and 10 stages, respectively (blue squares in **Fig. 2A**). Finally, for membranes with 3 μm pore size, the distillate mass flow rates were $J$ = 0.484±0.024, 1.443±0.068, and 2.610±0.043 L $m^{-2}$ $h^{-1}$ with 1, 3 and 10 stages, respectively (black triangles in **Fig. 2A**). Note that our 1-stage distiller presented distillate flow rates well below the thermodynamic limit of single-stage distillers under 900 W $m^{-2}$, that is 1.32 L $m^{-2}$ $h^{-1}$. In fact, the vapour transport occurring in traditional single-stage solar stills is typically dictated by (faster) convective processes rather than diffusive ones.

In these experiments, the ambient temperature was set to ≅ 22 °C, while the temperature differences measured across the distillers ranged from 54 °C at the top side of the evaporator to 45 °C at the bottom side of the condenser in the 1-stage setup (see **Supplementary Fig. S4A**). In the 3-stage setup, the temperature decreases from 60 °C at the top side of the stage one evaporator to 41 °C at the bottom side of the stage three condenser (see **Supplementary Fig. S4B**). Finally, in the 10-stage setup the temperature decreases from 60 °C at the top side of the stage one evaporator to 31 °C at the bottom side of the stage ten condenser (see **Fig. 2B**). In all cases, the temperature uncertainty was ±0.25 °C.

**Supplementary Note 2. Sensitivity analyses by theoretical model.** The passive distiller discussed in this work can treat in principle a broad variety of feed water types, spanning from low-salinity brackish water to high-salinity brines. Therefore, we modelled the effect of different feed water salinities on overall desalination performance. To study this in detail, we hypothesized a distiller with an evaporator-condenser interface including both a 0.2 mm air gap and a hydrophobic membrane in series, for treating feed water with NaCl concentrations of 35, 70 and 170 g $L^{-1}$. This hybrid interface was adopted to counterbalance the effect of high salinity on the partial pressure (equation (1) in the Methods) and thus allow the treatment of a broader range of water salinities. The modelling predictions reported in **Supplementary Fig. S10** clearly show that the distiller is more efficient with a lower feed water salinity; most importantly, this analysis indicates that the distiller can be used even with highly concentrated brines. [1] Furthermore, the results highlight that the distillate flow rate scales rather linearly with the number of stages up to a threshold, which depends on the feed water salinity. Beyond this threshold, the enhancement of distiller performance slows down because of a reduced temperature difference across each distillation stage, which is gradually less effective at counteracting the difference in vapour pressure caused by salinity.

Note that the productivity of the distiller could be possibly further enhanced by exploiting a concentrated solar source.

**Supplementary Note 3. Estimation of the energy required for rinsing.** The salt evacuation from the hydrophilic layers used as evaporators could be enhanced by rinsing them with a proper flow of saltwater (salinity: 35 g $L^{-1}$). This saltwater flow should be driven by a small hydraulic head ($\Delta z$ in **Fig. 3A**). In the durability experiments reported in **Fig. 3**, the saltwater basin needed for the rinsing process has been filled by hand. On the other side, the rinsing process could be also automatically carried out by a pump, whose electric power could be estimated as: $P_P \approx \frac{(g\Delta z)\rho\dot{V}}{\eta_P} = 3 \times 10^{-3}$ W, where $\rho = 1029$ kg m$^{-3}$ is the saltwater density, $\dot{V} = 2.5$ mL s$^{-1}$ is the volumetric flow rate required for rinsing (as tested in the experiments reported in **Fig. 3B**), $\Delta z = \Delta z_0 + \Delta z_1$ is the hydraulic head ($\approx 10$ cm), and $\eta_P = 0.8$ is the pump efficiency. Note that, due to the low flow velocities, both pressure losses and kinetic contributions are safely neglected. As in the considered experiments, the rinse pump should operate about $\Delta t = 60$ s per day, therefore $E_P = P_P\Delta t = 5 \times 10^{-5}$ Wh energy would be theoretically consumed per day for rinsing the evaporators in a 3-stage distiller. Note also that this is only an illustrative calculation far from any practical issues. To keep the desalination systems as completely solar-powered, a photovoltaic panel could be introduced to supply the estimated electric energy for the rinse pump. Considering a panel efficiency equal to $\eta_{PV}$ =12% and an average daily sum of global irradiation per square meter received by the PV modules equal to $H_d = 4.73$ kWh m$^{-2}$ (Torino, PVGIS database http://re.jrc.ec.europa.eu/pvgis/), a hypothetical $A_{PV} = 1$ cm$^2$ PV area would suffice to supply $E_{PV} = \eta_{PV} H_d A_{PV} \approx 5 \times 10^{-2}$ Wh. Therefore, a limited extension of photovoltaic panels (*i.e.*, 0.7% of the solar absorbing surface required by the distiller) would be largely ($\frac{E_{PV}}{E_P} \approx 1000$) sufficient to power the daily rinsing process.

**Supplementary Note 4. Estimation of the case study in Rome.** According to PVGIS database (http://re.jrc.ec.europa.eu/pvgis/), Rome has an average irradiation on horizontal plane equal to 7.75 kWh m$^{-2}$ day$^{-1}$ in July. By considering the configuration with 0.1 μm PTFE membrane and 10 stages tested under laboratory conditions, the measured distillate productivity rate of the distiller is 3.27 L kWh$^{-1}$. Since the daily freshwater need is 240 L per person per day in average and the current (2017) number of Rome inhabitants is 2875472, the total freshwater demand of Rome can be estimated as 690113 $m^3$ $day^{-1}$. By conservatively considering that only 50% of the average daily irradiation can be exploited by the distiller (e.g. due to the startup of evaporation process), the surface required to meet the total freshwater demand of Rome can be estimated as 54 $km^2$, that is a square with 7.4 km side.

## Supplementary Figures

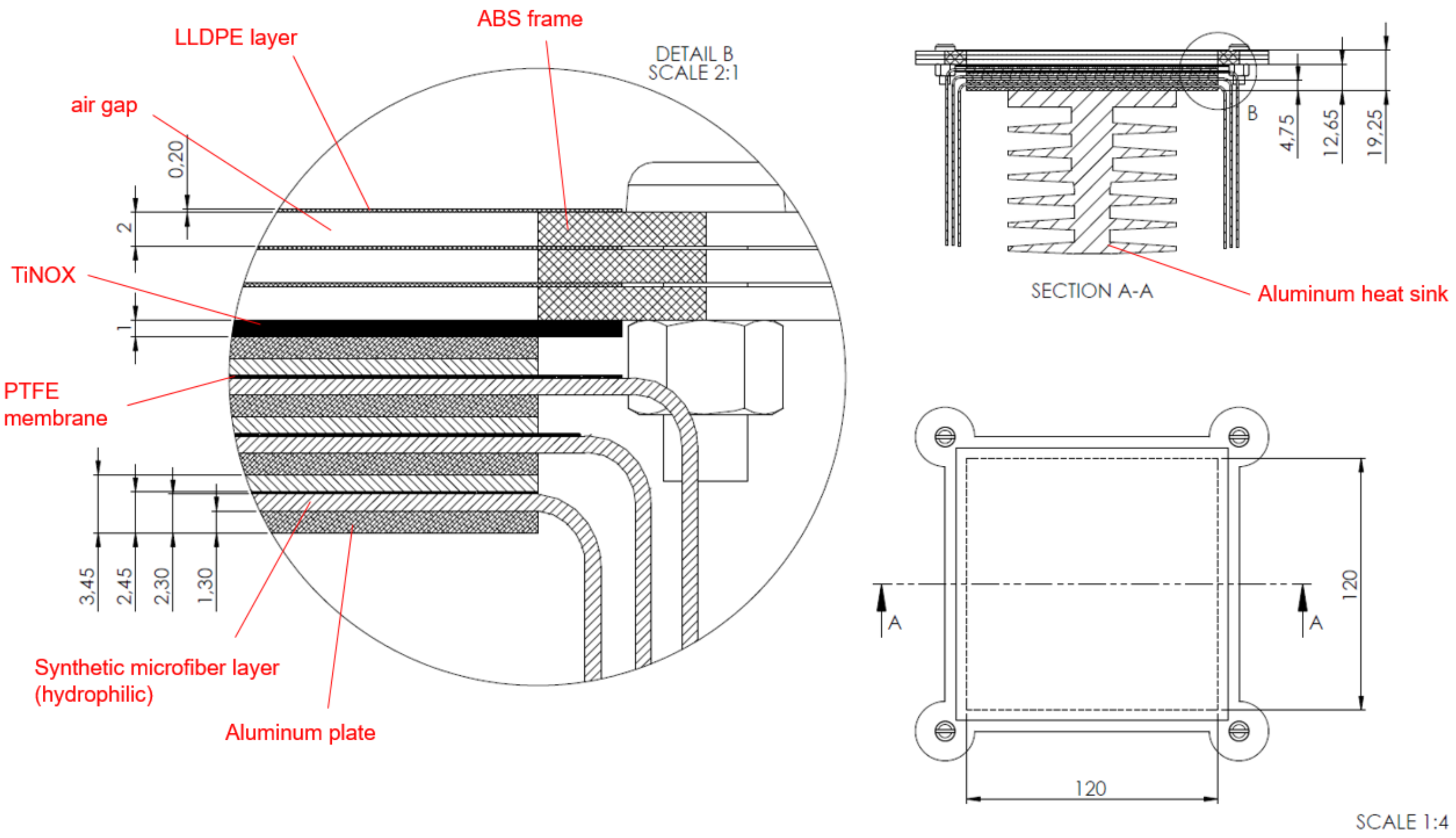

**Supplementary Figure S1. Technical draw of the tested prototype.** Top view and A-A section of the 3-stage solar distiller tested under laboratory and outdoor conditions (scale 1:4). In the top view, the reported dimensions refer to the solar collection area of the TiNOX® solar absorber. Detail B (scale 2:1) shows: the main parts of the thermal insulator for minimizing convective heat loss (ABS frame, LLDPE layers, and air gaps); the TiNOX® solar absorber; the assembly of layers per each distillation stage (synthetic microfiber, PTFE membrane, aluminium plate). The layers of the distillation stages are stacked one on each other, and the stability of the overall assembly is ensured by a few pieces of tape applied to the lateral sides of stages. Note that the 1-stage and 10-stage configurations of the distiller are implemented by simply removing or adding distillation stages. Dimensions are reported in millimetres.

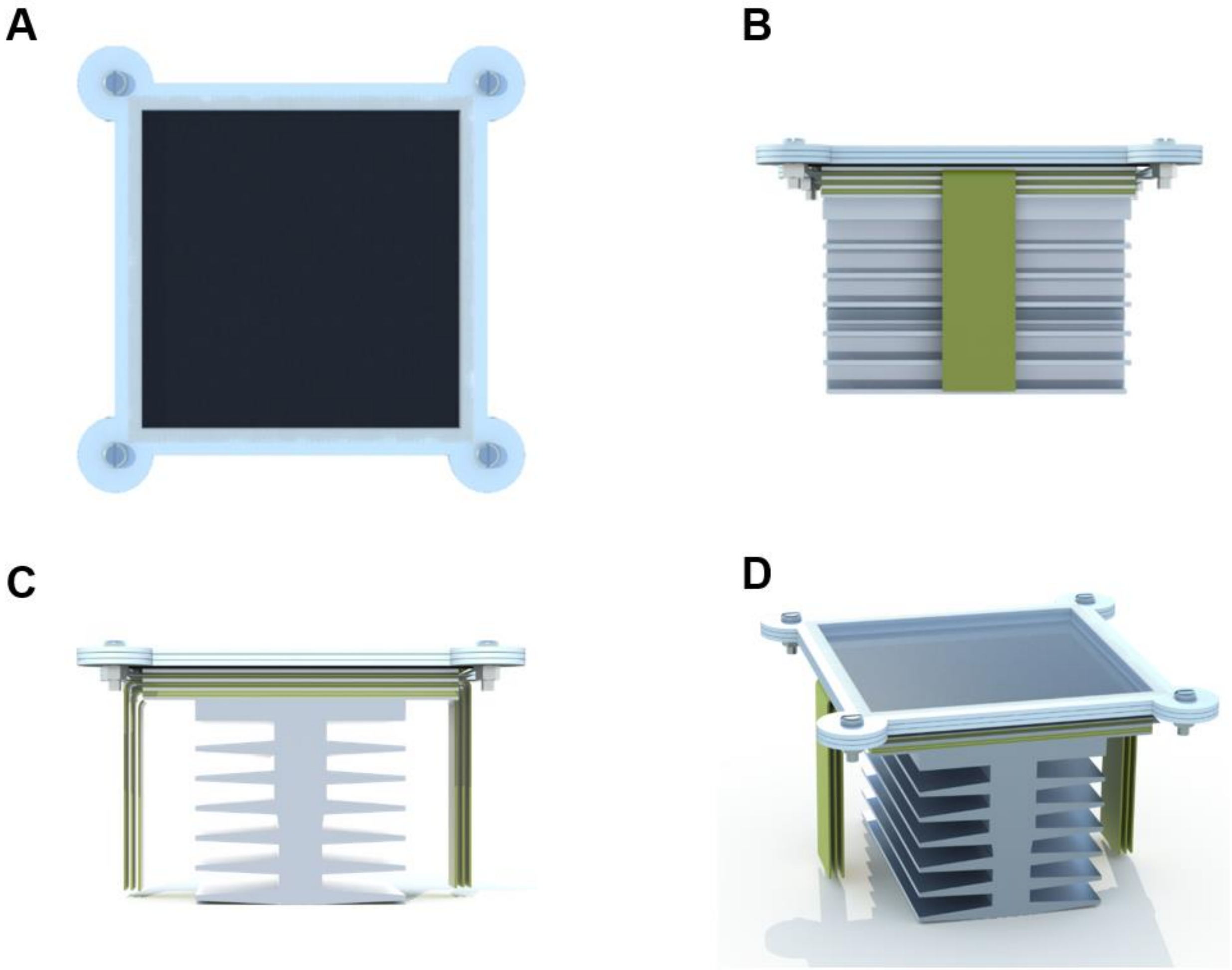

**Supplementary Figure S2. 3D model of the solar desalination prototype. (A)** Top view. **(B)** Lateral view. **(C)** Front view. **(D)** Isometric view. Note that the yellow, hydrophilic strips are then immersed in either saltwater (evaporators) or distilled (condensers) water basins. Dimensions of components are reported in **Supplementary Fig. S1**.

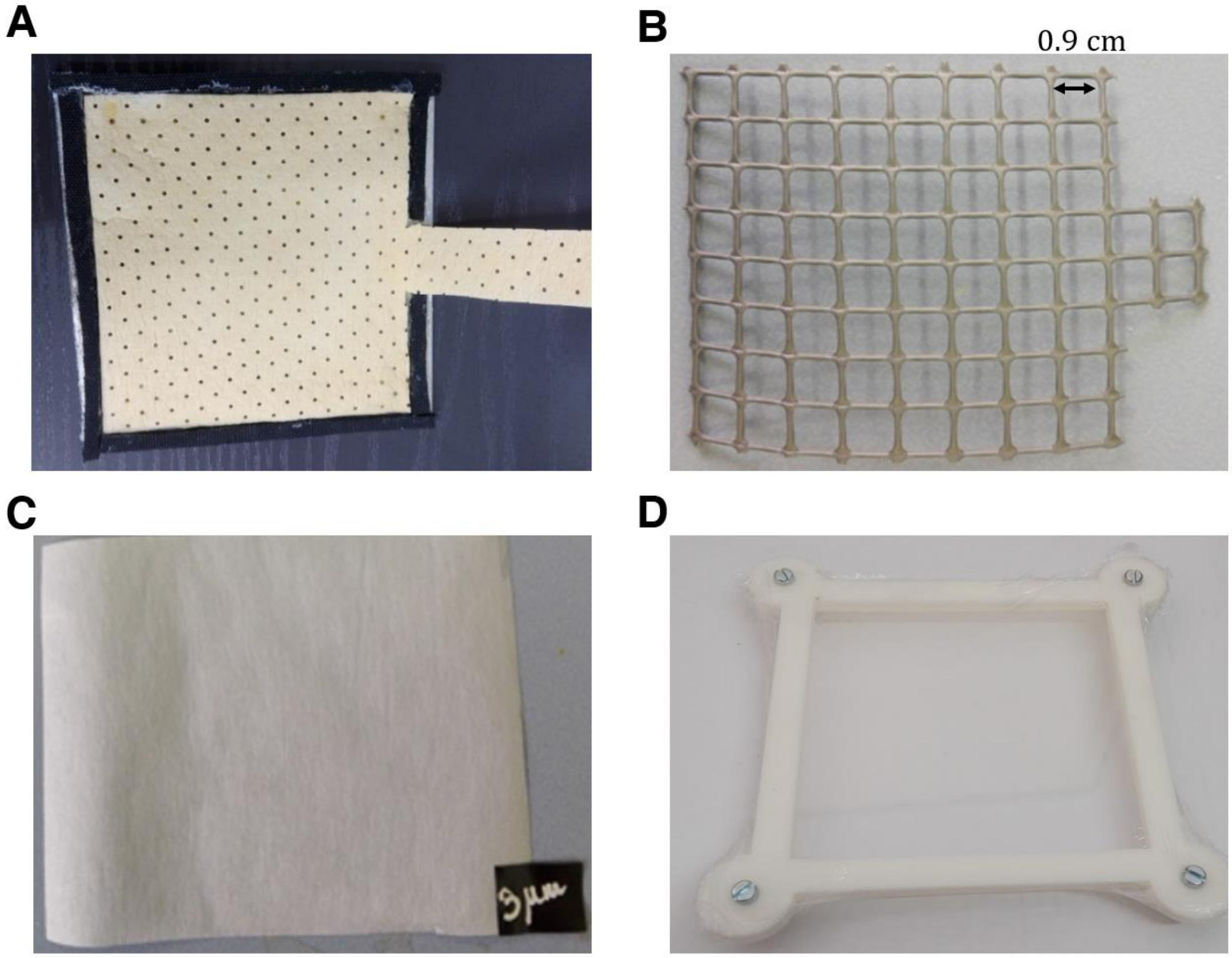

**Supplementary Figure S3. Main components of the solar desalination prototype. (A)** Hydrophilic layer (synthetic microfiber made of 70% viscose fibre, 18% polypropylene, and 12% polyester) for passive seawater supply or distilled water discharge. **(B)** Spacer made of polypropylene for air gap creation (membrane-free configuration). Note that the separation between hydrophilic layers could be also achieved by embedding a rigid frame or mesh between the metallic plates. Combinations of both hydrophobic membranes and air gaps are also possible. **(C)** Microporous membrane with pore size 3 μm made of polytetrafluoroethylene (*ANOW Microfiltration co., LTD*). Note that also the membranes with 0.1 μm pore size are made of polytetrafluoroethylene (*ANOW Microfiltration co., LTD*). **(D)** Thermal insulator placed above the distiller, to minimize convective heat loss. This component is made of an ABS frame manufactured by 3D printing (Fused Deposition Modelling, *Stratasys Elite*) and three parallel layers of transparent LLDPE.

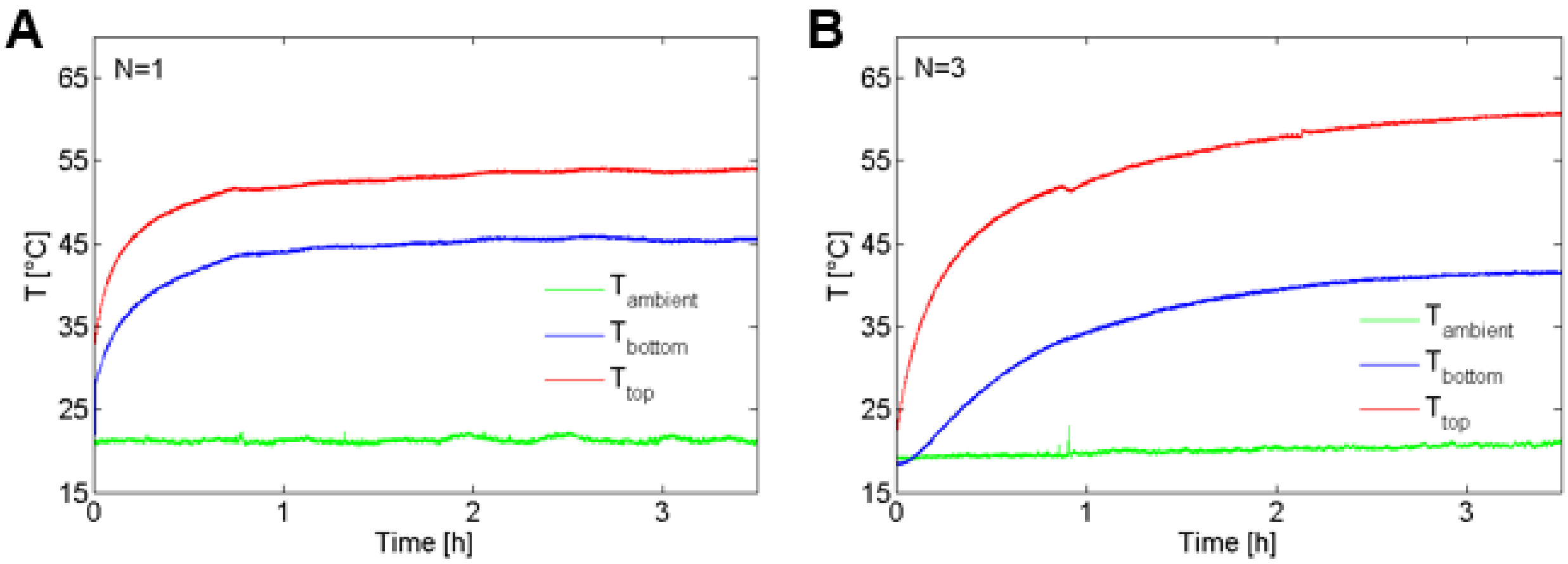

**Supplementary Figure S4. Temperature profiles in the distiller (laboratory tests).** The temperature profiles refer to the laboratory tests of different configurations of the modular distiller. The tested input thermal power is equal to 740 W $m^{-2}$, which corresponds to an equivalent solar irradiance approximately equal to 900 W $m^{-2}$. **(A)** 1-stage configuration. **(B)** 3-stage configuration. Red, blue, and green lines represent the first stage evaporator (top side of distiller), last stage condenser (bottom side of distiller) and ambient temperatures, respectively. Note that the temperature gradient between condenser and evaporator is about 10 °C in the 1-stage setup, 20 °C in the 3-stage setup, and 30 °C in the 10-stage setup (see **Fig. 2B**).

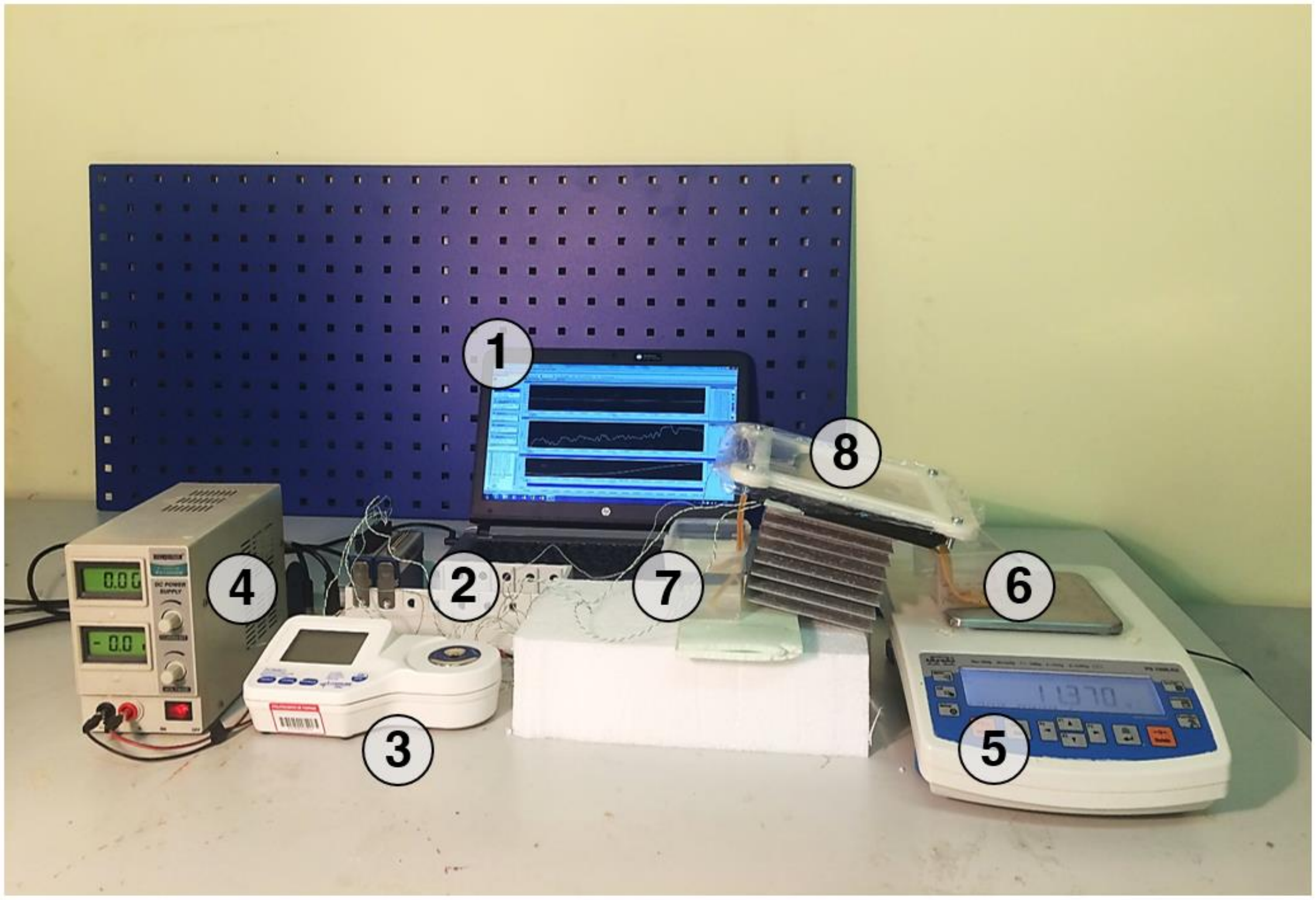

**Supplementary Figure S5. Laboratory testing of the modular distiller.** The following components were present in the experimental tests carried out under laboratory conditions: (1) a laptop for data storage and elaboration, (2) a data acquisition board, (3) a digital refractometer for salinity measurement, (4) a power supply unit, (5) a scale for distillate mass measurement, (6) an output basin for the distilled water, (7) an input basin for the saltwater, and (8) the modular distiller.

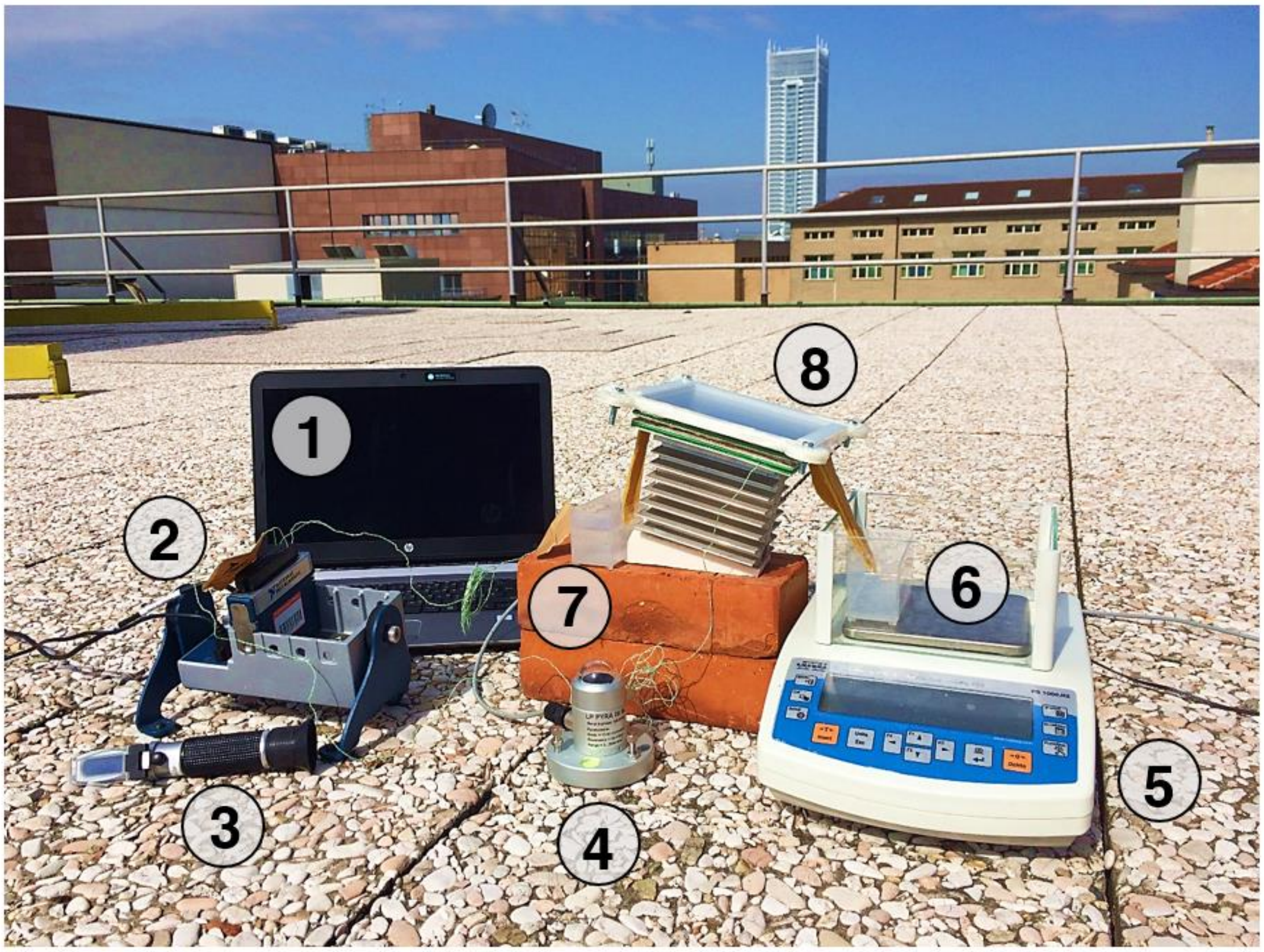

**Supplementary Figure S6. Outdoor (roof) testing of the modular distiller.** The following components were present in the experimental tests carried out the roof of Department of Energy (Politecnico di Torino) under outdoor conditions: (1) a laptop for data storage and elaboration, (2) a data acquisition board, (3) an analog refractometer for salinity measurement, (4) a pyranometer for solar irradiance measurement, (5) a scale for distillate mass measurement, (6) an output basin for the distilled water, (7) an input basin for the saltwater, and (8) the modular distiller. The experimental tests were performed in Torino, Piedmont, Italy.

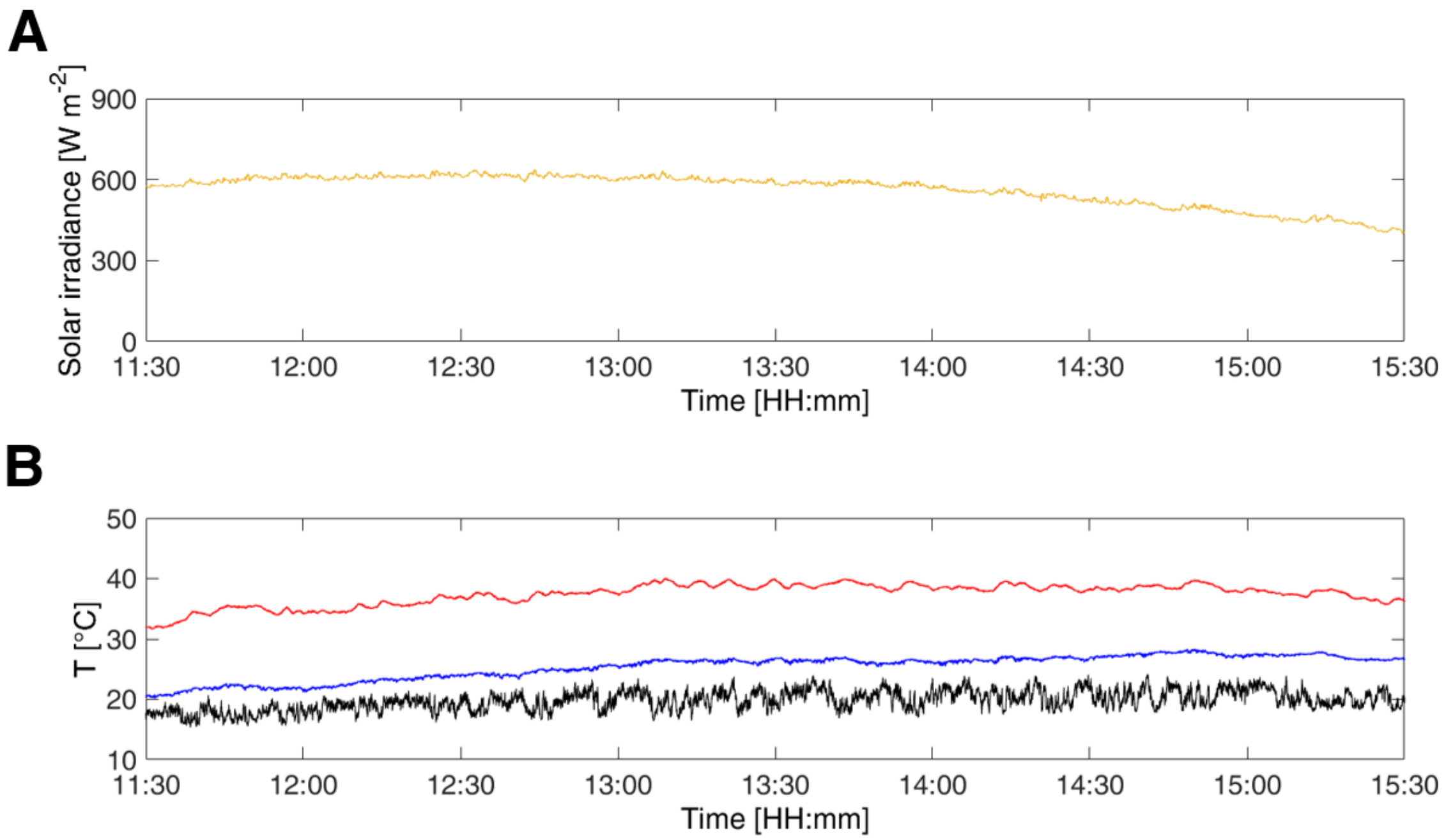

**Supplementary Figure S7. Outdoor (roof) conditions of the 3-stage distiller. (A)** Solar irradiance (16 March 2017, 45°03'43.0"N 7°39'35.7"E Torino - Italy) and **(B)** temperature profiles during outdoor tests. The distillate flow rate and its uncertainty has been computed from approximately 1 pm to 2:30 pm. Red, blue, and black lines represent the first stage evaporator (top side of distiller), last stage condenser (bottom side of distiller), and ambient temperatures, respectively. During the experiment, the average temperature difference measured across the distiller ranged from 38 °C for the stage one evaporator to 26 °C for the stage three condenser.

A B

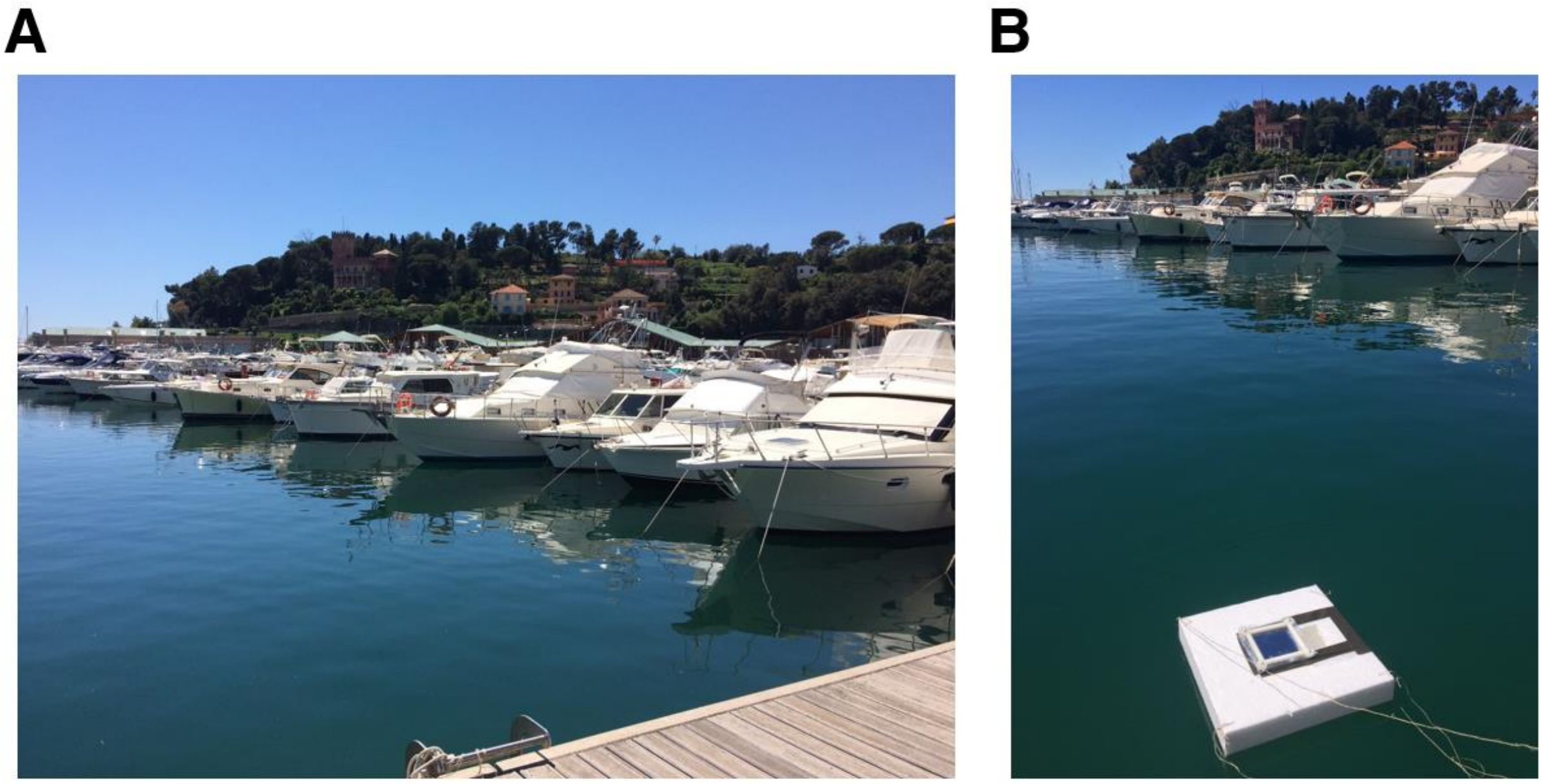

**Supplementary Figure S8. Outdoor test of the 3-stage distiller floating above the Ligurian Sea. (A)** Overview of the experimental environment (Marina di Varazze, Varazze - Italy) and **(B)** floating 3-stage distiller during outdoor test.

A | B

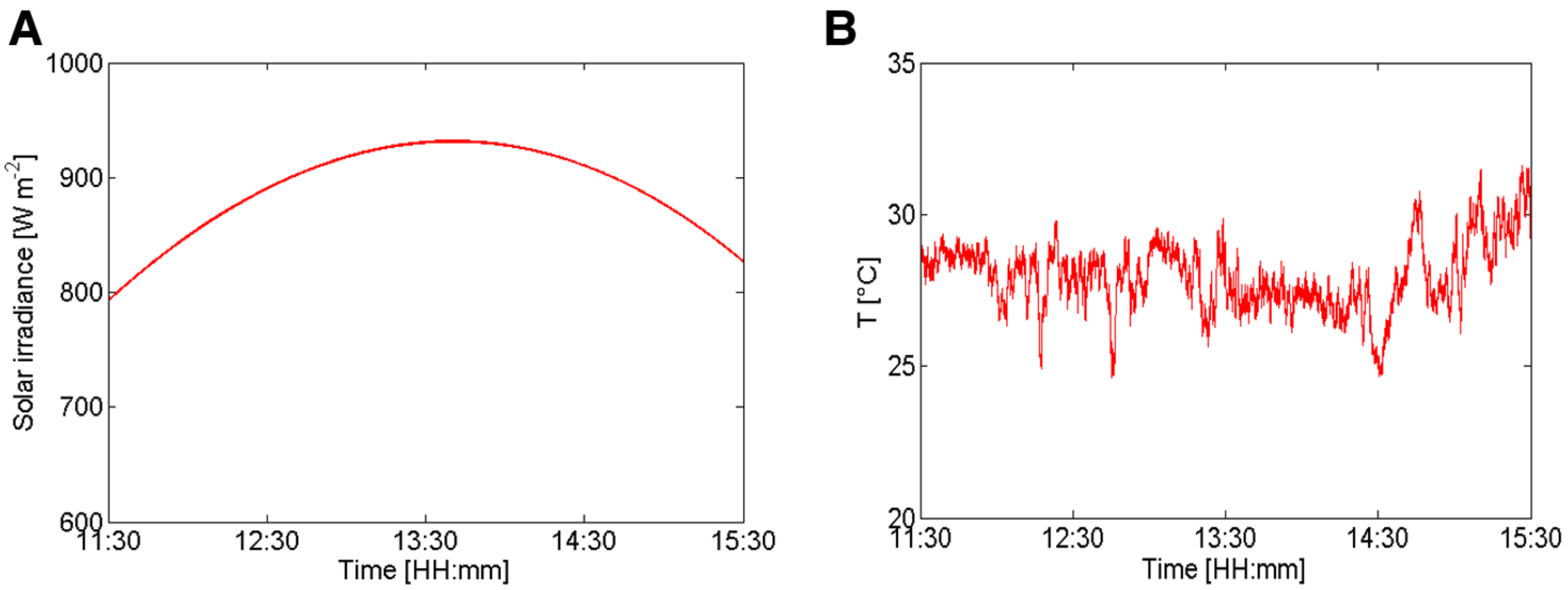

**Supplementary Figure S9. Ambient data during outdoor (sea) tests of the floating 3-stage distiller. (A)** Solar irradiance (17 May 2017, 44°21'14.3"N 8°33'58.0"E Varazze - Italy) and **(B)** ambient temperature during outdoor test. The distillate flow rate has been computed from approximately 11:30 am to 3:30 pm.

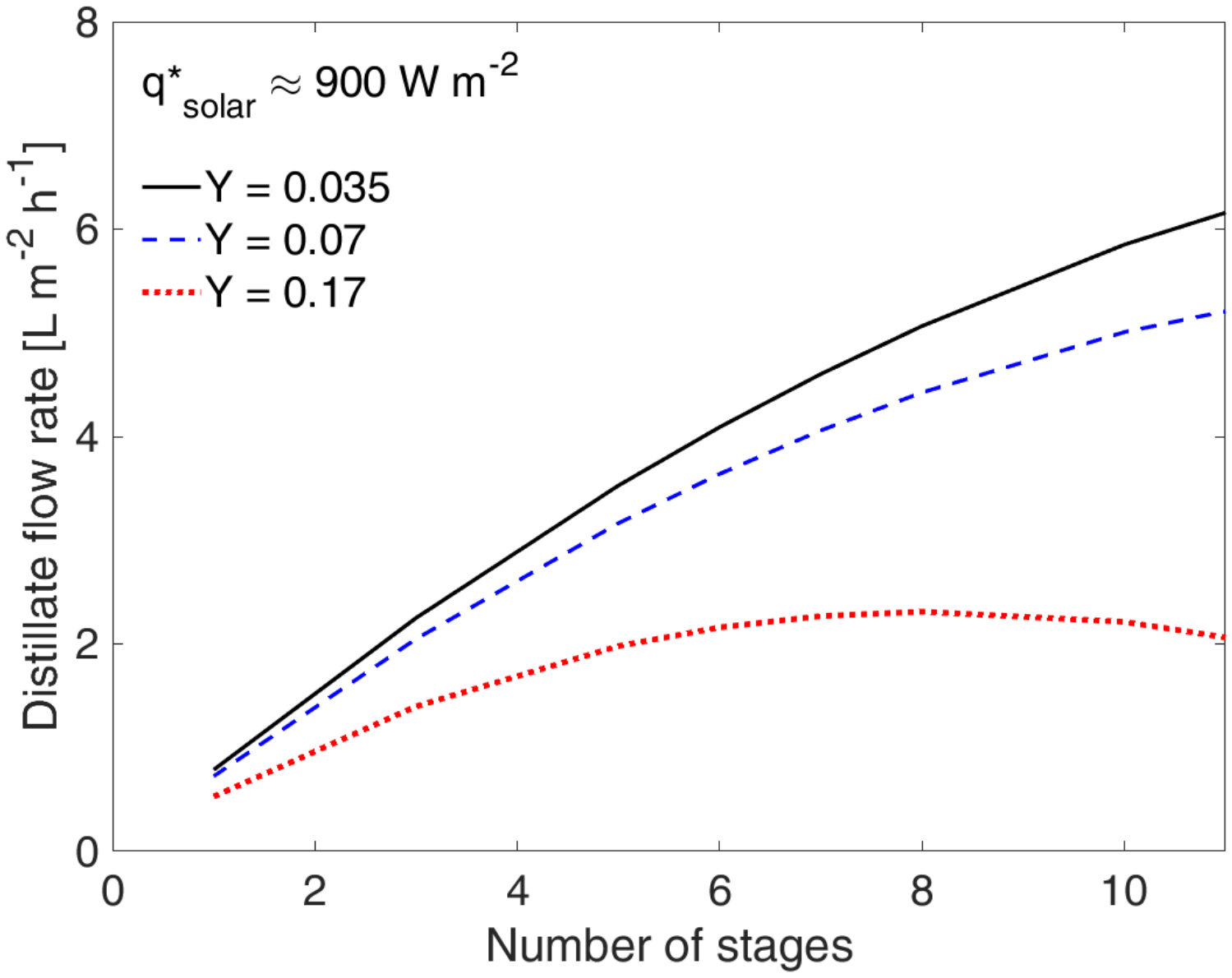

**Supplementary Figure S10. Modeling estimates of optimized distillation performances.** Potential desalination performances of the modular distiller using input saltwater with different NaCl concentrations, as predicted by the theoretical model in equations (1-7): 35 g $L^{-1}$ ($Y = 0.035$), 70 g $L^{-1}$ ($Y = 0.070$) and 170 g $L^{-1}$ ($Y = 0.170$) are represented. In the model, the considered equivalent solar irradiance was 900 W $m^{-2}$, and the interface between the evaporation and condensation layers includes both a 0.2 mm air gap and a hydrophobic membrane with 3.0 μm pore size (see **Supplementary Note 2** for details).

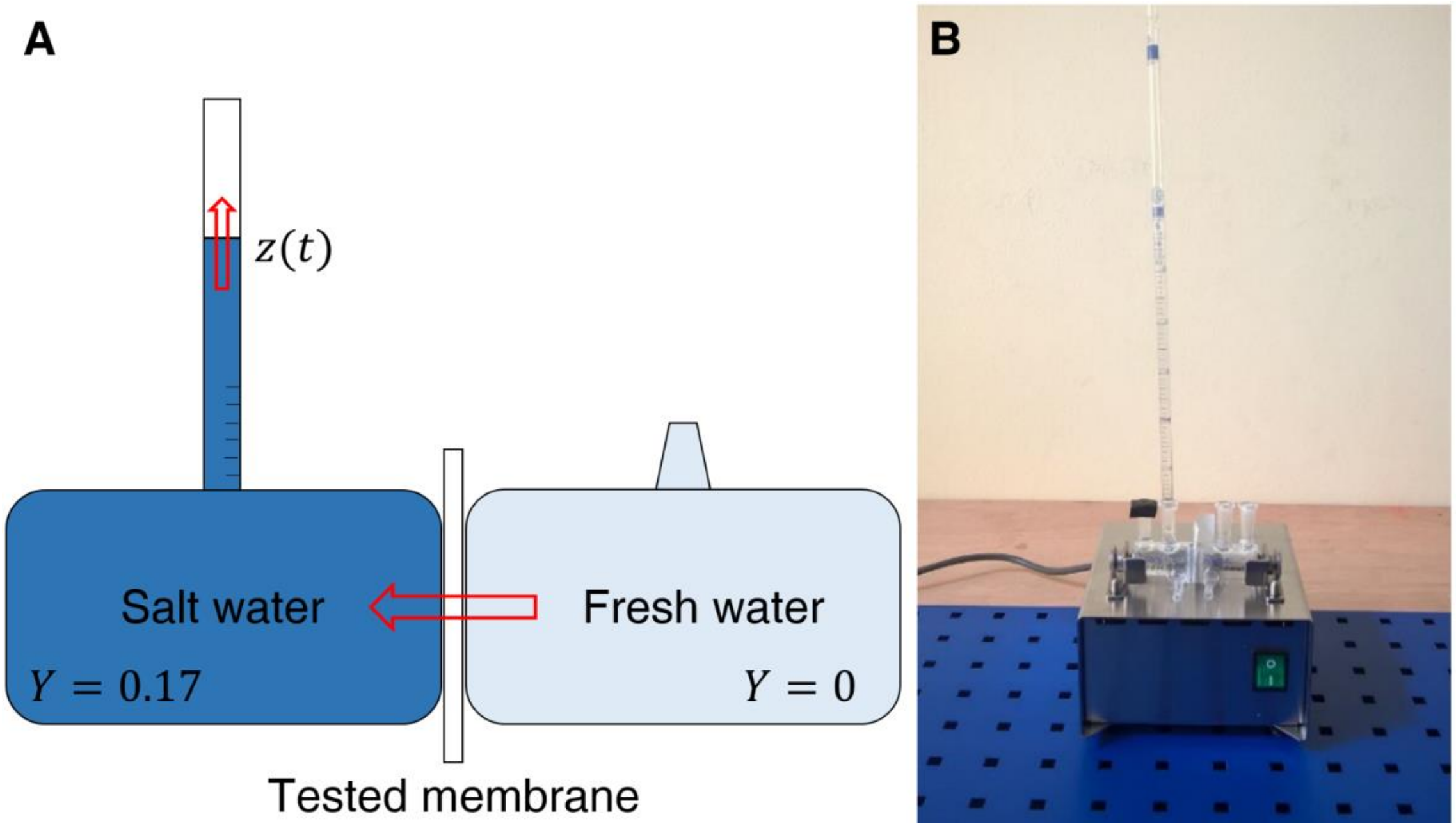

**Supplementary Figure S11. Experimental measure of membrane permeability. (A)** Schematics and **(B)** picture of the diffusion cell employed to assess the permeability of a membrane clamped between water/NaCl and distilled water solutions.

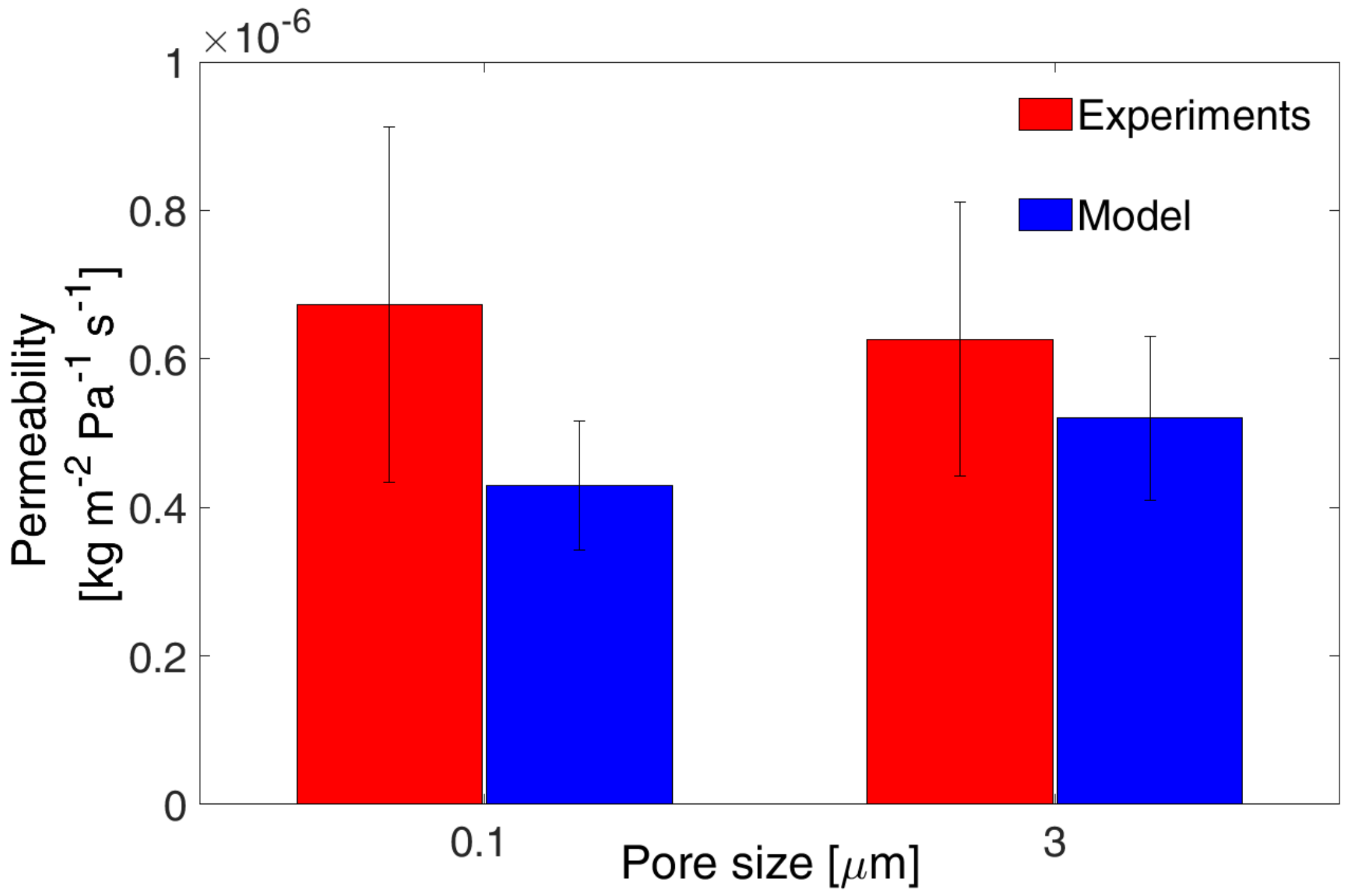

**Supplementary Figure S12. Experimental and predicted membrane permeability.** Membrane permeability measured by the experimental setup shown in **Supplementary Fig. S11** are compared with modeling estimations (equations (5) and (6), considering 21°C temperature and 1 bar pressure). PTFE membranes with either 0.1 μm or 3.0 μm average pore size are considered (*ANOW Microfiltration co., LTD*).

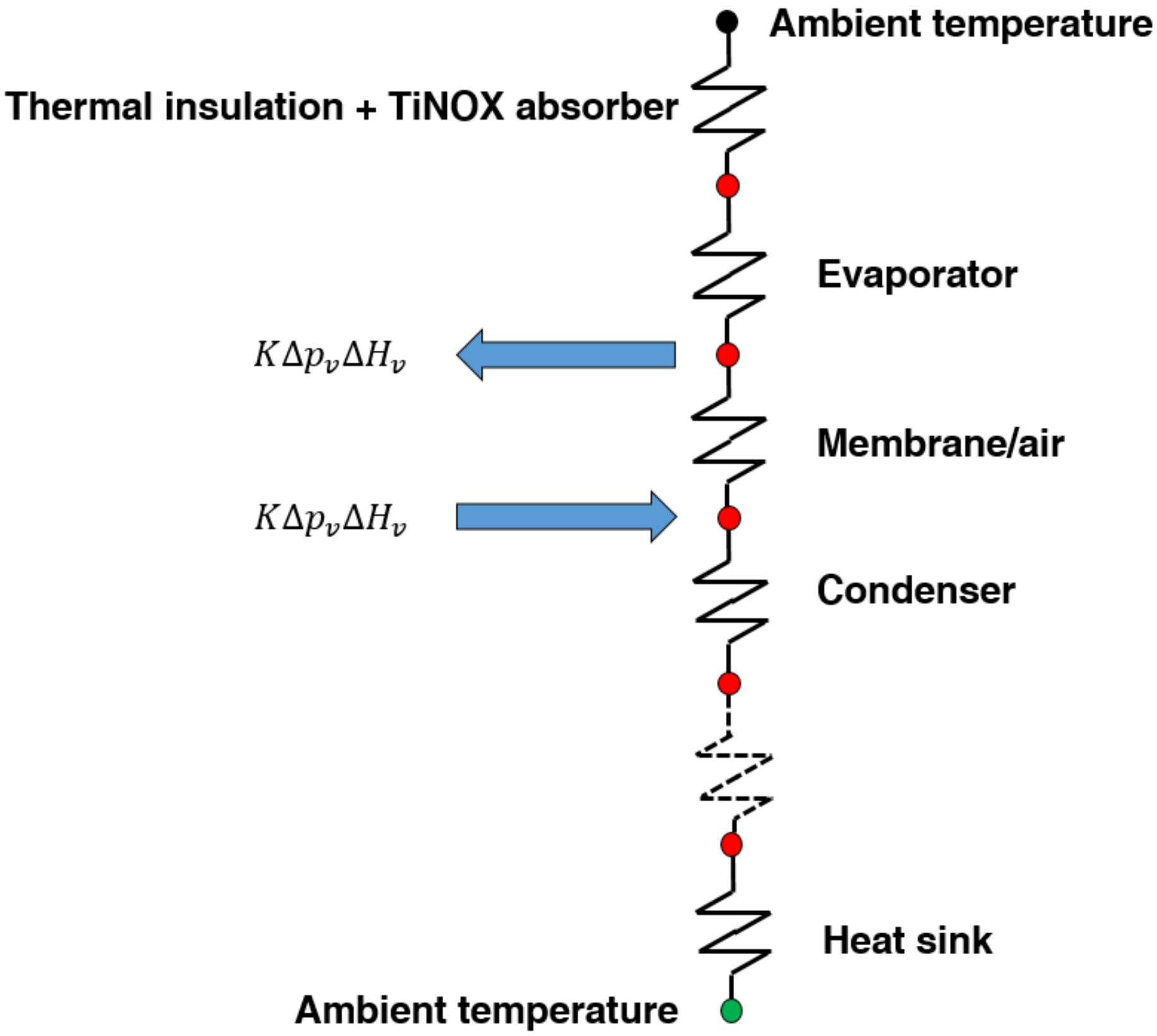

**Supplementary Figure S13. Heat conduction through the distiller.** Lumped thermal resistances considered in the one-dimensional model to interpret the heat conduction through the distiller. The thermal resistances are due to (from top to bottom): convective and conductive heat transfer between the first stage evaporator and ambient, through the TiNOX® solar absorber and the convective heat transfer insulator; series of thermal conduction resistances due to the multiple layers in the N stages of the distiller; convective heat transfer between heat sink and ambient. Note that latent heat transfer due to evaporation/condensation of water in the distiller is also modelled (blue arrows).

## SUPPLEMENTARY TABLES

| Heat source | Test location | Interface type | N. of stages | $q_{solar}$ [W m$^{-2}$] | J [L m$^{-2}$ h$^{-1}$] | $e_s$ [kWh m$^{-3}$] | $1/e_s$ [L kWh$^{-1}$] |
|---|---|---|---|---|---|---|---|
| Electrical resistor | Laboratory | Membrane-free | 1 | 900±9 | 0.498±0.023 | 1807±85 | 0.553±0.026 |
| Electrical resistor | Laboratory | 0.1 µm PTFE membrane | 1 | 900±9 | 0.481±0.007 | 1871±33 | 0.534±0.009 |
| Electrical resistor | Laboratory | 3.0 µm PTFE membrane | 1 | 900±9 | 0.484±0.024 | 1860±94 | 0.538±0.027 |
| Electrical resistor | Laboratory | Membrane-free | 3 | 900±9 | 1.462±0.050 | 616±22 | 1.624±0.058 |
| Electrical resistor | Laboratory | 0.1 µm PTFE membrane | 3 | 900±9 | 1.442±0.040 | 624±18 | 1.602±0.047 |
| Electrical resistor | Laboratory | 3.0 µm PTFE membrane | 3 | 900±9 | 1.443±0.068 | 624±30 | 1.603±0.077 |
| Electrical resistor | Laboratory | Membrane-free | 10 | 900±9 | 2.892±0.124 | 311±14 | 3.213±0.142 |
| Electrical resistor | Laboratory | 0.1 µm PTFE membrane | 10 | 900±9 | 2.947±0.024 | 305±4 | 3.274±0.042 |
| Electrical resistor | Laboratory | 3.0 µm PTFE membrane | 10 | 900±9 | 2.610±0.043 | 345±7 | 2.900±0.056 |
| Sun | Roof | 3.0 µm PTFE membrane | 3 | 593±11 | 0.585±0.038 | 1014±68 | 0.987±0.067 |
| Sun | Roof | 3.0 µm PTFE membrane | 10 | 637±20 | 1.320±0.055 | 483±25 | 2.072±0.108 |
| Sun | Sea | 3.0 µm PTFE membrane | 3 | 850 | 1.500 | 567 | 1.765 |

**Supplementary Table S1. List of experimental results.** List of desalination performances of the passive distillers tested in this work with different heat source, test location, evaporator/condenser interface, and number of distillation stages. In the table, the reported quantities are: $q_{solar}$, average solar – or solar equivalent – irradiance during tests; $J$, average mass flow rate of produced distilled water per square meter exposed to the solar radiation; $e_s$, solar energy required to produce a cubic meter of distilled water; $1/e_s$, litres of distilled water produced by 1 kWh of input solar energy. Average values and standard deviations are both reported.

| Experimental setup | $h$ [W m$^{-2}$ K$^{-1}$] | | $\varepsilon$ [-] | | $\delta$ [μm] | | $T_{sky}$ [K] |
|---|---|---|---|---|---|---|---|
| | Lower bound | Upper bound | Lower bound | Upper bound | Lower bound | Upper bound | |
| Laboratory | 5 | 7 | 0.75 | 0.86 | 5 | 10 | 295 |
| Roof | 7 | 10 | 0.75 | 0.86 | 5 | 10 | 263 |
| Sea surface | 7 | 10 | 0.75 | 0.86 | 5 | 10 | 263 |

**Supplementary Table S2. Uncertainties in the theoretical model.** Upper and lower values of the variables in the theoretical model have been adopted to determine the uncertainty of the model estimations: $h$, convective heat transfer coefficient (distiller-ambient); $\varepsilon$, porosity of the hydrophobic membrane; $\delta$, additional air gap between hydrophilic layers and hydrophobic membrane due to non-ideal contacts; $T_{sky}$, sky temperature to estimate radiative heat losses. Upper and lower bounds have been inferred from experimental evidences.

| Category | $q_{solar}$ [W m$^{-2}$] | S [m$^2$] | J [L m$^{-2}$ h$^{-1}$] | $e_s$ [kWh m$^{-3}$] | $1/e_s$ [L kWh$^{-1}$] | Type | Reference | Note |
|---|---|---|---|---|---|---|---|---|
| 3 stages, PTFE 0.1 µm | 900 | 1.44E-02 | 1.442 | 624 | 1.602 | Passive | This work | Experiments |
| 10 stages, PTFE 0.1 µm | 900 | 1.44E-02 | 2.947 | 305 | 3.274 | Passive | This work | Experiments |
| 15 stages, PTFE 0.1 µm | 900 | 1.44E-02 | 3.920 | 230 | 4.356 | Passive | This work | Estimations |
| Advanced solar steam | 1000 | 1.96E-03 | 1.076 | 929 | 1.076 | Passive | Ref. 2 | Experiments |
| Advanced solar steam | 2000 | 1.96E-03 | 1.850 | 1081 | 0.925 | Passive | Ref. 2 | Experiments |
| Advanced solar steam | 3000 | 1.96E-03 | 2.960 | 1014 | 0.987 | Passive | Ref. 2 | Experiments |
| Advanced solar steam | 5000 | 1.96E-03 | 5.280 | 947 | 1.056 | Passive | Ref. 2 | Experiments |
| Advanced solar steam | 6600 | 1.96E-03 | 7.670 | 860 | 1.162 | Passive | Ref. 2 | Experiments |
| Advanced solar steam | 8500 | 1.96E-03 | 9.950 | 854 | 1.171 | Passive | Ref. 2 | Experiments |
| Advanced solar steam | 9000 | 1.96E-03 | 10.790 | 834 | 1.199 | Passive | Ref. 2 | Experiments |
| Advanced solar steam | 10000 | 1.96E-03 | 12.000 | 833 | 1.200 | Passive | Ref. 2 | Experiments |
| Advanced solar steam | 1000 | 3.14E-04 | 0.708 | 1412 | 0.708 | Passive | Ref. 3 | Experiments |
| Advanced solar steam | 2000 | 3.14E-04 | 1.200 | 1667 | 0.600 | Passive | Ref. 3 | Experiments |
| Advanced solar steam | 4000 | 3.14E-04 | 2.370 | 1688 | 0.593 | Passive | Ref. 3 | Experiments |
| Advanced solar steam | 6000 | 3.14E-04 | 3.750 | 1600 | 0.625 | Passive | Ref. 3 | Experiments |
| Advanced solar steam | 10000 | 3.14E-04 | 6.670 | 1499 | 0.667 | Passive | Ref. 3 | Experiments |
| Advanced solar steam | 1000 | 6.16E-04 | 0.950 | 1053 | 0.950 | Passive | Ref. 4 | Experiments |
| Advanced solar steam | 2000 | 6.16E-04 | 2.220 | 901 | 1.110 | Passive | Ref. 4 | Experiments |
| Advanced solar steam | 3000 | 6.16E-04 | 3.670 | 817 | 1.223 | Passive | Ref. 4 | Experiments |
| Advanced solar steam | 4000 | 6.16E-04 | 5.640 | 709 | 1.410 | Passive | Ref. 4 | Experiments |
| Advanced solar steam | 6000 | 6.16E-04 | 8.610 | 697 | 1.435 | Passive | Ref. 4 | Experiments |
| Low-cost solar steam | 1000 | 8.61E-04 | 1.100 | 909 | 1.100 | Passive | Ref. 5 | Experiments |
| Low-cost solar steam | 10000 | 1.00E-04 | 11.800 | 847 | 1.180 | Passive | Ref. 6 | Experiments |
| Low-cost solar steam | 1000 | 5.31E-04 | 1.450 | 690 | 1.450 | Passive | Ref. 7 | Experiments |
| Low-cost solar steam | 1000 | 1.60E-03 | 1.280 | 781 | 1.280 | Passive | Ref. 8 | Experiments |
| Low-cost solar steam | 3000 | 1.60E-03 | 3.660 | 820 | 1.220 | Passive | Ref. 8 | Experiments |
| Low-cost solar steam | 5000 | 1.60E-03 | 6.240 | 801 | 1.248 | Passive | Ref. 8 | Experiments |
| Low-cost solar steam | 7000 | 1.60E-03 | 9.340 | 749 | 1.334 | Passive | Ref. 8 | Experiments |
| Low-cost solar steam | 10000 | 1.60E-03 | 13.300 | 752 | 1.330 | Passive | Ref. 8 | Experiments |
| Low-cost solar steam | 12000 | 1.00E-04 | 14.020 | 856 | 1.168 | Passive | Ref. 9 | Experiments |
| Low-cost solar steam | 1000 | 7.85E-03 | 0.518 | 1931 | 0.518 | Passive | Ref. 10 | Experiments |
| Direct irradiation + MD | 700 | 2.80E-03 | 0.300 | 2333 | 0.429 | Passive | Ref. 11 | Experiments |
| Direct irradiation + MD | 700 | 2.80E-03 | 0.100 | 7000 | 0.143 | Passive | Ref. 11 | Experiments |
| Direct irradiation + MD | 700 | 1 | 0.500 | 1400 | 0.714 | Passive | Ref. 11 | Estimations |
| Direct irradiation + MD | 700 | 1 | 0.200 | 3500 | 0.286 | Passive | Ref. 11 | Estimations |
| Solar still | 441 | 3.76E-01 | 0.412 | 1070 | 0.935 | Passive | Ref. 12 | Experiments |
| Solar still | 622 | 4.90E-02 | 0.392 | 1585 | 0.631 | Passive | Ref. 13 | Experiments |
| Solar still | 850 | 6.40E-01 | 0.700 | 1214 | 0.824 | Passive | Ref. 14 | Experiments |
| Solar still | 700 | 6.80E-01 | 0.936 | 748 | 1.337 | Passive | Ref. 15 | Experiments |
| Solar still - combined | 600 | 5.00E-01 | 1.530 | 392 | 2.550 | Active | Ref. 16 | Experiments |
| Solar still - combined | 650 | 3.14E+00 | 0.344 | 1889 | 0.530 | Active | Ref. 17 | Experiments |
| Solar still - combined | 520 | 9.60E-02 | 2.083 | 250 | 4.006 | Active | Ref. 18 | Experiments |
| Solar still - combined | 500 | 1.00E+00 | 0.625 | 800 | 1.250 | Active | Ref. 19 | Simulations |
| Solar collectors + MD | 500 | 2.00E+04 | NA | 3869 | 0.258 | Active | Ref. 20 | Estimations |
| Solar collectors + MD | 500 | 1.90E+04 | NA | 3648 | 0.274 | Active | Ref. 20 | Estimations |
| Solar collectors + MD | 500 | 4.00E+03 | NA | 750 | 1.334 | Active | Ref. 20 | Estimations |
| Solar collectors + MD | 550 | 1.80E+01 | 0.711 | 773 | 1.293 | Active | Ref. 21 | Experiments |
| Solar collectors + MD | 588 | 2.00E+01 | 7.840 | 75 | 13.333 | Active | Ref. 22 | Experiments |
| Solar collectors + MD | 588 | 2.00E+01 | 2.940 | 200 | 5.000 | Active | Ref. 22 | Experiments |
| Solar collectors + MD | 1000 | 7.00E+00 | 2.143 | 467 | 2.143 | Active | Ref. 23 | Experiments |
| Solar collectors + MD | 950 | 5.73E+00 | 4.363 | 218 | 4.593 | Active | Ref. 24 | Experiments |
| Solar collectors + MD | 600 | 2.16E+00 | 0.460 | 1304 | 0.767 | Active | Ref. 25 | Experiments |
| PV + RO | 330 | NA | NA | 22 | 45.000 | Active | Ref. 26 | Experiments |
| PV + RO | 500 | NA | NA | 43 | 23.286 | Active | Ref. 27 | Estimations |

**Supplementary Table S3. Performances of solar desalination technologies.** Comparison of the energy performances of several passive or active solar desalination technologies. Passive technologies: current work (experimental laboratory results of 3- and 10-stage prototypes with PTFE membranes, 0.1 μm pores; estimations of 15-stage prototype with PTFE membranes, 0.1

μm pores); advanced materials for solar steam generation; membrane distillation (MD) process powered by direct solar irradiation; low-cost materials for solar steam generation; solar stills. Active technologies: solar stills combined with active components (e.g. pumps); membrane distillation (MD) process powered by solar thermal collectors; reverse osmosis (RO) process powered by photovoltaic (PV) panels. In the table, the reported quantities are: $q_{solar}$, average solar – or solar equivalent – irradiance during tests; $S$, surface exposed to solar radiation; $J$, average mass flow rate of produced distilled water per square meter exposed to the solar radiation; $e_s$, solar energy required to produce a cubic meter of distilled water; $1/e_s$, litres of distilled water produced by 1 kWh of input solar energy.

## SUPPLEMENTARY REFERENCES